\documentclass{aa}  
\usepackage{natbib}
\bibpunct{(}{)}{;}{a}{}{,} % to follow the A&A style

\usepackage{graphicx}	% Including figure files
\usepackage{txfonts}

\usepackage{amsmath}	% Advanced maths commands
\usepackage{amssymb}	% Extra maths symbols
\usepackage{import}
\usepackage{multirow}
\usepackage{subfigure}

\newcommand{\fpath}{figure/subfigure}
\newcommand{\bvec}[1]{\mathbf{#1}}

\begin{document}

   \title{Numerical study of Kelvin-Helmholtz instability and its impact on synthetic emission from magnetized jets}

  % \subtitle{I. Overviewing the $\kappa$-mechanism}

   \author{Nikhil Borse\thanks{E-mail: nb29100@gmail.com}
          \inst{1} \and
          Sriyasriti Acharya\inst{1} \and Bhargav Vaidya\inst{1} \and Dipanjan Mukherjee\inst{2} \and 
          Gianluigi Bodo\inst{3} \and Paola Rossi\inst{3} \and Andrea Mignone\inst{4}
          }

   \institute{Discipline of Astronomy Astrophysics and Space Engineering, Indian Institute of Technology Indore, Khandwa Road, Simrol, Indore 453552, India \and 
   Inter-University Centre for Astronomy and Astrophysics, Post Bag 4, Pune - 411007, India
\and INAF/Osservatorio Astrofisico di Torino, Strada Osservatorio 20, I-10025 Pino Torinese, Italy
\and Dipartimento di Fisica Generale, Universita degli Studi di Torino , Via Pietro Giuria 1, 10125 Torino, Italy }

   \date{Received XXX; accepted XXX}

% \abstract{}{}{}{}{} 
% 5 {} token are mandatory
 
  \abstract
   {Non-thermal emission from Active Galactic Nuclei (AGN) jets extends up-to large scales in-spite of them being prone to a slew of magneto-hydrodynamic instabilities.} {The main focus of this study is to understand the impact of MHD instabilities on the non-thermal emission from large scale AGN jets.} {We perform high resolution three-dimensional numerical magneto-hydrodynamic simulations of a plasma column to investigate the dynamical and emission properties of jet configurations at kilo-parsec scales with different magnetic field profiles, jet speeds, and density contrast. We also obtain synthetic non-thermal emission signatures for different viewing angles using an approach that assumes static particle spectra and that obtained by evolving the particle spectra using Lagrangian macro-particles incorporating the effects of shock acceleration and radiative losses.} {We find that the shocks due to Kelvin-Helmholtz (KH) instability in the axial magnetic field configurations can strongly affect the jet dynamics. Additionally, we also find the presence of weak biconical shocks in the under-dense jet columns. The inclusion of a helical magnetic field hinders the vortex growth at the shear surface thereby stabilizing the jet column. With the evolving particle spectra approach, the synthetic SEDs obtained for cases with strong KH instability show the presence of multiple humps ranging from radio to TeV gamma-ray band.} {We conclude that the high energy electrons accelerated in the vicinity of freshly formed shocks due to KH instability, result in high X-ray emission.}


   \maketitle
%
%-------------------------------------------------------------------

\section{Introduction}

Jets from Active Galactic Nuclei represent channels through which energy is transported from the central black hole region,  where they are formed and accelerated, towards the extended regions, up to scales that can reach the mega-parsec, where the jet terminates.  Along their way,  part of their energy can be dissipated and transferred to a relativistic electron population, giving rise to the observed non-thermal radiation, which can cover the complete electromagnetic spectrum, i.e., from radio up to $\gamma$-rays. In this process, jet instabilities play a fundamental role since they initiate the dynamical processes that lead to energy dissipation. Two kinds of instabilities are mainly considered in this context: Kelvin-Helmholtz instabilities (KHI), driven by the velocity shear between the jet and the ambient medium, and current-driven instabilities (CDI), that develop in the presence of a helical magnetic field. KHI leads to the formation of shocks and turbulence through which the energy is dissipated. In the case of CDI instead, dissipation occurs mainly through the formation of current sheets and magnetic reconnection.
   
Much literature has focused on understanding the characteristics of these instabilities in  MHD jets, determining the physical parameters that control their growth. KHI have been extensively studied in several different configurations, both in Newtonian and in relativistic jets and with or without the presence of a magnetic field both performing a linear analysis \citep[see e.g.][]{Bodo_1989, Birkinshaw_1991, Hardee_1992, Bodo_1996, Kersale2000, Urpin_2002,  Perucho_2004, Perucho_2010, 2006AIPC..856...57H, 2007ApJ...662..835M} and following the nonlinear evolution through numerical simulations \citep{Frank:1996, Malagoli1996, Ryu2000}. The main parameters that control the behavior of the instability are the Mach number,  the density ratio between the jet and ambient medium, and the plasma $\beta$ in the presence of a magnetic field.  The magnetic field configuration also has a big impact on the instability evolution; a longitudinal field may have a stabilizing effect, while a toroidal field may introduce the other class of instabilities discussed above, i.e., CDI. Linear analysis of CDI has been performed both in the Newtonian and in the relativistic cases by performing linear analysis \citep[see e.g.][]{Appl1992, Istomin94, Istomin96, Bodo:2013, Kim15, Bodo16, Kim16, Kim17, Kim18, Bodo19} and following the evolution by numerical simulations.  When both instabilities are possible \citet{Appl1992} showed that KHI are stabilized, and this result was confirmed by \citet{Baty_2002} through three-dimensional numerical simulations.  \citet{Rossi_2008, Mukherjee_2020} showed that instabilities might have also a big impact on the large scale morphological properties of jets, leading either to the transition to an FRI type morphology (see also \cite{2016A&A...596A..12M}) or, in higher power jets, to strongly different properties of the cocoon.
   
Here we are interested in another effect of instabilities; as discussed above, the instability evolution leads to energy dissipation and consequently to the formation of a non-thermal electron population and the observed emission. Connecting all these processes is a formidable challenge since they involve a huge range of physical scales, where emission processes occur at the micro-scale, while the dynamics happen at the macro-scale. For this reason, studies of jets have concentrated either on the dynamics, sometimes providing emission properties by using very simplified recipes, or on the radiation, employing very simplified dynamical structures.  Few studies have tried more sophisticated approaches, in which the evolution of a non-thermal electron population subject to energy losses and gains is followed together with the dynamics of the thermal fluid \citep{Micono1999, Tregillis:2001, 2013JPhCS.454a2001M, Vaidya_2018, Fromm:2019, Winner_2019, Huber:2021}, under the assumption that non-thermal particles are advected with the fluid velocity and have no back-reaction on the fluid.  In particular, energy gains at shocks are treated in a sub-grid manner by employing more or less sophisticated recipes for determining the properties of the non-thermal particle distribution function depending on the shock characteristics. 
In particular, we have adopted the treatment for the particle acceleration process at shocks developed within the hybrid particle framework for the PLUTO code \citep{Vaidya_2018}.
   
The main goal of this work is to analyze the possible imprint of KHI on the jet radiation properties, as it was done in \citep{Micono1999} and, since our focus is on the instability evolution, we will adopt an idealized approach, where we consider a small section of a jet, assuming that in the nearby regions all jet sections behave in the same way. With this approach, we can employ an adequate numerical resolution to accurately follow the evolution of KHI. Our focus will be on the jet properties on a scale of tens of kpc; at these distances, the jet has most likely slowed down from the highly relativistic regime at its base, reaching non-relativistic or mildly relativistic velocities. For this reason, our simulations are performed by using the non-relativistic MHD equations.
   
The paper is organized as follows. Section 2 describes the required numerical methodology and initial conditions for dynamical modeling. In section 3, different emission modelling approaches are presented. Section 4 describes the results obtained from dynamical analysis and emission modeling with the effects of orientation. The impact of shock formation, particle acceleration, and instabilities on the emission signatures are explained in section 5. Finally, section 6 lists the important findings of this work.

%--------------------------------------------------------------------

\section{Numerical Setup}\label{setup}

\subsection{Equations and numerical methodology}

Three-dimensional simulations of cylindrical plasma columns have been carried out by solving the following set of ideal time-dependent magneto-hydrodynamic equations in the Cartesian coordinate system (X, Y, Z) -  
\begin{equation}
    \frac{\partial \rho}{\partial t} + \nabla \cdot (\rho \bvec{v}) = 0,
\end{equation}
\begin{equation}
        \frac{\partial \rho \bvec{v}}{\partial t} + \nabla \cdot \left(\rho \bvec{v} \bvec{v}  - \bvec{B} \bvec{B} + \left(P + \frac{\bvec{B} \cdot\bvec{B}}{2} \right) \bvec{I} \right) = 0,
\end{equation}
\begin{equation}
        \frac{\partial E}{\partial t} + \nabla \cdot \left[\left(P + E +  \frac{B^{2}}{2} \right)\bvec{v} -\bvec{B}(\bvec{v} \cdot \bvec{B}) \right] = 0,
\end{equation}
\begin{equation}
        \frac{\partial \bvec{B}}{\partial t} + \nabla \cdot [ \bvec{v} \bvec{B}  - \bvec{B} \bvec{v} ] = 0,
\end{equation}
\begin{equation}
    \nabla \cdot \bvec{B} = 0,
\end{equation}
where $\rho$, $P$, $\bvec{B}$, and $\bvec{v}$ are the density, isotropic gas pressure, magnetic field, and velocity respectively. Note that a factor $1/\sqrt{4\pi}$ has been reabsorbed in the definition of $\bvec{B}$. The energy density is the sum of thermal, kinetic, and magnetic energy densities respectively. It is given by the following expression 
\begin{equation}
    E = \frac{P}{\Gamma -1} + \frac{\rho v^{2}}{2} + \frac{B^{2}}{2},
\end{equation}
where the internal energy is governed by the ideal gas equation and the ratio of specific heats, $\Gamma$ is 5/3. Further, we employ a scalar tracer field $\tau$ to distinguish between the jet and the ambient fluid. Its value is set to unity for the region $r < R_{\rm j}$, where $r = \sqrt{X^2 + Y^2}$ is the cylindrical radius and $R_{\rm j}$ is the jet radius.

The numerical simulations are carried out employing the MHD module of the astrophysical fluid dynamics code PLUTO \citep{Mignone_2007}. A linear reconstruction shock-capturing method employing the \textit{hllc} solver is used.

\subsection{Initial conditions}

We simulate a portion of an AGN jet at kpc scales assuming that the jet has become non-relativistic \citep{Laing2014}. 
For this purpose, a 3D cylindrical plasma column is initialized in a Cartesian box of size $ 4 R_{\rm j} \times 4 R_{\rm j} \times L_{\rm z}$. Here, $R_{\rm j}$ is the radius of the plasma column which is set to unity and $L_{\rm z}$ is the axial size of the box following an aspect ratio $L_{\rm z}/R_{\rm j} = 2$. The resolution of the grid is set to $ 200 \times 200 \times 100 $ zones which translates to $\Delta X = \Delta Y = \Delta Z = 0.02 R_{\rm j}$. We will refer to this 3D plasma column as a \textit{jet} but it should be noted that we are only modeling a representative section of the large-scale kpc jet as similar dynamical features are likely to also occur at other regions in this portion of the jet.
The simulation runs are typically done using non-dimensional quantities and expressed in code units (c.u.). 
To scale them in physical units, we have chosen three scales - the jet radius ($R_{j}$ = 100 pc), a reference velocity ($V_{\rm sc}$ = 5000 km s$^{-1}$), and the external density ($\rho_{\rm out} = 1.004 \times 10^{-26}$ g cm$^{-3}$) relevant for the present study. As a consequence, the unit for the magnetic field results to be $177 \mu$G. 

The density in the jet is set using a parameter $\eta = \rho_{\rm out}/\rho_{\rm 0}$ that represents the density contrast between the ambient value ($\rho_{\rm out} = 1$ ) and that on the jet axis $\rho(r = 0) = \rho_{\rm 0}$. We model jets with a density equal to ambient ($\eta = 1$) and also those which are under-dense ($\eta > 1$). The density profile is
\begin{equation}
    \rho(r) = \rho_{\rm out}\left(1 + \frac{(1/\eta - 1)}{\cosh{\left(\frac{r}{R_{\rm j}}\right)^4}}\right).
\label{eq:Equation7}
\end{equation}

The flow velocity is set along the axial direction $\bvec{\hat{z}}$ and sheared radially with a hyperbolic tangent profile. The initial radial distribution of the velocity field is given as \citep{Baty_2002}
\begin{equation}
    V_{\rm z}(r) = \frac{V}{2} \tanh \left( \frac{R_{\rm j} - r}{a} \right),
\end{equation}
\\
where $V$ is the amplitude of the velocity shear, $R_{\rm j}$ is the jet radius, a is the width of the shear layer and r is the radius in the cylindrical coordinate system.
The values of these initial model parameters are $R_{\rm j} = 1$, $a = 0.1$, and on-axis pressure $P(r=0)= P_{\rm 0} = 1$. The sonic speed on the jet axis  is given by $c_{\rm s} = (\Gamma P_{\rm 0}/ \rho_{\rm 0})^{1/2}$. 
For the jet models with a density equal to ambient ($\eta = 1$), the value of sound speed is $c_{\rm s} = 1.29$ which, in physical units, is 
6450 kms$^{-1}$ corresponding to a thermal temperature of $\sim 3 \times 10^{9} \,K$ at the axis of the jet. The flow regime by Mach number can be found by calculating the Mach number along the axis using $M_{\rm s} = V/c_{\rm s}$ where $V$ is the speed of the jet at the axis. 

The radial profile of the magnetic field structure can be expressed in the following general form
 given by \cite{Baty_2002}
\begin{equation}
    B_{\rm r} = 0 ,\qquad B_{\rm \phi} = B_{\rm 1} \frac{r/r_{\rm c}}{1 + (r/r_{\rm c})^{2}} ,\qquad B_{\rm z} = B_{\rm 0},
\label{eq:Equation9}
\end{equation}
\\
where the parameters $B_{\rm 0}$ and $B_{\rm 1}$ control the magnetic field strength and the radial pitch profile and $r_{\rm c}$ is the characteristic column radius. 

\begin{figure}
\centering
\includegraphics[width = 8.6cm, scale=1.0]{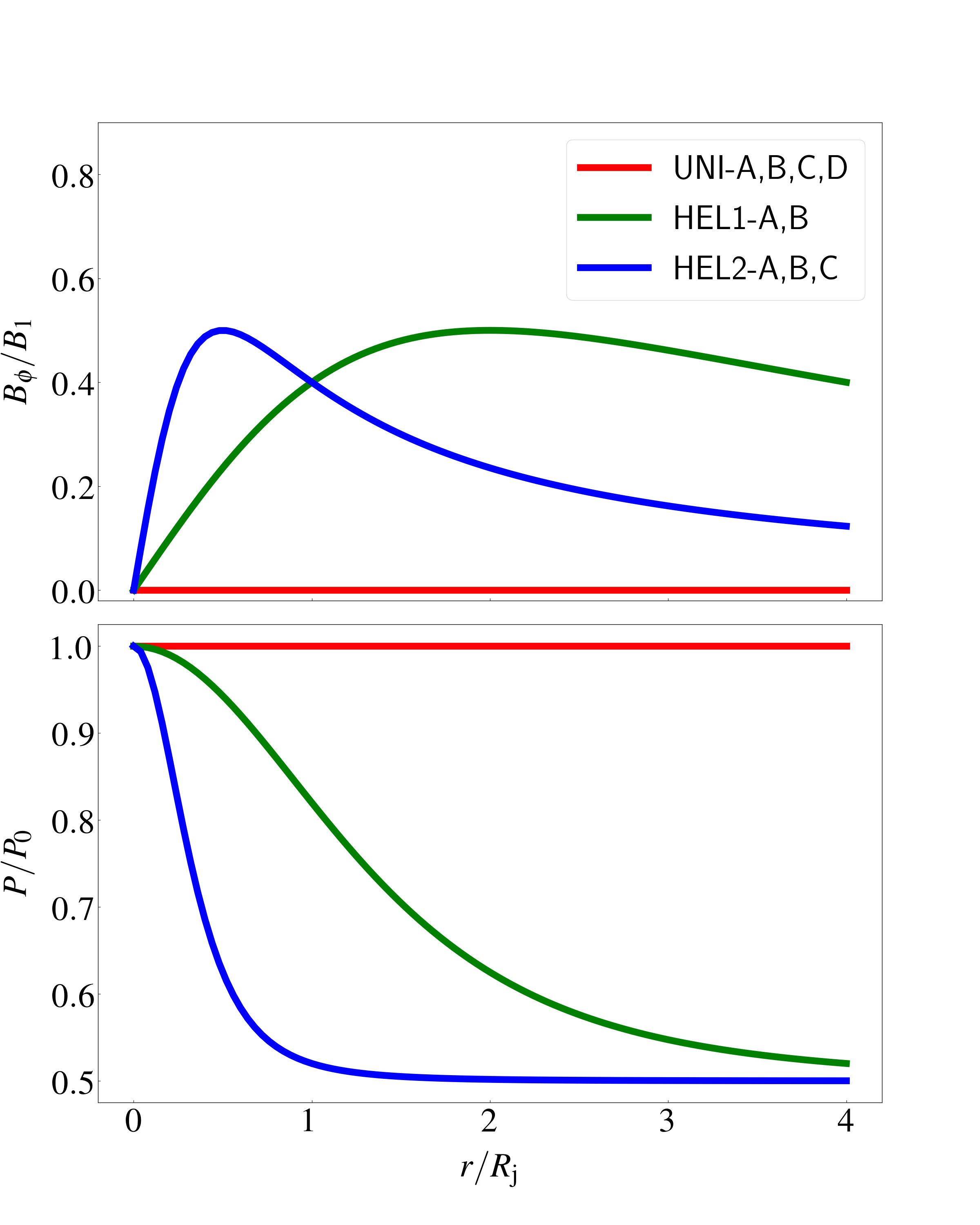}
\caption{Initial radial profiles of the azimuthal magnetic field $B_{\rm \phi}$ (top), and gas pressure $P$ (bottom) for all the jet models given in table~\ref{tab:Table1}. 
}
\label{fig:Figure1}
\end{figure}

For positive values of the parameter $B_{\rm 1}$, the current density has its maxima on the jet axis. The initial radial profiles of the azimuthal magnetic field $B_{\rm \phi}$ are shown in Fig.~\ref{fig:Figure1}. For the system to be in a state of magneto-hydrodynamic equilibrium initially, which is essentially a balance between gas pressure and magnetic pressure forces, the gas pressure distribution must follow the radial component of the momentum conservation equation which is given by 
\begin{equation}
    \nabla P = {(\nabla \times \bvec{B}) \times \bvec{B}}.
\end{equation}
The radial profile of the gas pressure distribution can then be derived analytically from the radial component of the above expression to yield the form
\begin{equation}
    P = P_{\rm 0} - \frac{{B_{\rm 1}}^{2}}{2 \rho_{\rm 0}} \left( 1 - \frac{1}{[1 + (r/r_{\rm c})^{2}]^{2}}
\right),
\label{eq:Equation13}
\end{equation}
\\
and is shown in Fig.~\ref{fig:Figure1} for all configurations.
We further define the radial pitch profile as follows - 
\begin{equation}
    \Pi (r)= \frac{r B_{\rm z}}{B_{\rm \phi}} = \frac{r_{\rm c} B_{\rm 0}}{B_{\rm 1}} \left[1 + \left(\frac{r}{r_{\rm c}} \right)^{2} \right],
\label{eq:pitch_prof}
\end{equation}
\\
where the choice of $r_{\rm c}$ governs the position of the maximum of pitch $\Pi$.

\begingroup
\begin{table*}
\centering
\caption{Summary of parameters in the initial configuration of the jet. Here, the magnetic pitch parameter $\Pi$, sonic Mach number $M_{\rm s}$, and axial velocity $V_{\rm z}/c$ are specified on the jet axis, $B_{\rm \phi}$ is the azimuthal magnetic field, $\eta$ is the density contrast, $t_{\rm stop}/t_{\rm a}$ is the time at which the run ends, and $r_{\rm c}$ is the characteristic column radius. All the cases have plasma $\beta = 2 P_{0}/B_{0}^{2} = 32$ on the jet axis and an axial magnetic field $B_{\rm z}$ = 44 $\mu$\,G. }
\label{tab:Table1}
\setlength{\tabcolsep}{10pt} % Default value: 6pt
\renewcommand{\arraystretch}{1.5} % Default value: 1
\begin{tabular}{|c|c|c|c|c|c|c|c|c|}\hline\hline
Case & $r_{\rm c}/R_{\rm j}$ & $\Pi(r = 0)/R_{\rm j}$ & $M_{\rm s}(r = 0)$ &  $V_{\rm z}(r = 0)/c$ & $B_{\rm \phi}(r = R_{\rm j})$ & $\eta$ & $t_{\rm stop}/t_{\rm a}$ & Remarks \\   
& & & &$\mu$\,G & & & & \\  \hline 
UNI-A & ... & $\infty$  &  1.26  &  0.027 & 0 & 1 & 3.75 &  \\ 
HEL1-A & 2 & 0.5 &  1.26  &  0.027  & 71 & 1 & 3.75 & Cases for validation \\ 
HEL2-A & 0.5 & 0.125 &  1.26  &  0.027 & 71 & 1 & 3.75 & \\ \hline 
UNI-B & ... & $\infty$  &  5.00  &  0.1 & 0 & 1 & 4.75 & Reference case \\  \hline 
HEL1-B & 2 & 0.5 &  5.00  &  0.1 & 71  & 1 & 4.75 &  \\ 
HEL2-B & 0.5 & 0.125 &  5.00  &  0.1 & 71  & 1 & 4.75 & \\ 
UNI-C & ... & $\infty$  &  5.00  &  0.24 & 0 & 5 & 7.25 & Comparative cases\\ 
HEL2-C & 0.5 & 0.125 &  10.00  &  0.2 & 71 & 1 & 4.75 & \\ 
UNI-D & ... & $\infty$  &  5.00  &  0.7 & 0 & 50 & 15.0 & \\  \hline 
\end{tabular}
\end{table*}
\endgroup

A periodic boundary condition is used along the axial direction which restricts the wavelength of the perturbations to values that fit within the length $L_{\rm z}$, which is the size of the computational domain along the axial direction. The boundary condition on the side walls is chosen as reflective to have a helical field structure in the jet.

The magneto-hydrodynamic equilibrium is perturbed only using the $ m = \pm 1$ modes. The mathematical form of the velocity perturbations applied to the three cases is given by

\begin{equation}
    v_{\rm r} = \delta V \exp\left(-\frac{(r - R_{\rm j})^2}{16a^2} \right) \cos (m \theta) \sin \left(\frac{2\pi n Z}{L_{\rm z}} \right),
\label{eq:Equation15}
\end{equation}
\\
where $\delta V = 0.01 $ is the amplitude of the applied velocity perturbation and $ 4a = 0.2$ is its width along the radial direction. The axial and azimuthal numbers are set to $n=1$ and $m=1$ to excite the Fourier modes that may play a role in destabilizing the jet. Such perturbations can trigger Kelvin-Helmholtz modes on the surface and kink modes in the presence of non-zero current density as there is a small yet finite displacement of the jet localized at the boundary or the jet radius $R_{\rm j}$.

We have studied a total of nine models with varying sonic Mach number $M_{\rm s}$, density contrast $\eta$, characteristic column radius $r_{\rm c}$, and magnitude of axial and azimuthal component of magnetic field strength. A summary of the detailed model parameters for all the cases is given in Table~\ref{tab:Table1}.
The initial conditions for the jet models with $M_{\rm s} = 1.26$ (i.e., the $A$ cases) follow that of \cite{Baty_2002}. The $A$ cases are primarily for validation purposes. Based on the magnetic field structure, the jet models are classified as the UNI, HEL1 and HEL2 cases and are described as follows.

\begin{itemize}
   
    \item The UNI cases have a uniform axial magnetic field. The model UNI-B is our reference case. In addition to UNI-B, we investigate the dynamical evolution of the model UNI-A for validation and under-dense jets UNI-C and UNI-D for comparison. The UNI-B, C, and D cases have an axial sonic Mach number $M_{\rm s} = 5.0$ while the UNI-A case has  $M_{\rm s} = 1.26$. The UNI-C and D cases are under-dense jets that have a smooth density variation along the radial direction. The UNI-D configuration has the highest density contrast $\eta$.
    
    \item The HEL1 cases are current carrying magnetized jets with an initial helical magnetic field having a pitch profile with characteristic column radius $r_{\rm c} = 2.0$ (see Eq.~\ref{eq:pitch_prof}). The HEL1-B case is for comparison with the reference case UNI-B and the HEL1-A case is for validation. The HEL1-A and B cases have the same axial sonic Mach numbers of $M_{\rm s} = 1.26$ and $M_{\rm s} = 5.0$ respectively as their UNI counterparts. 
    
    \item The HEL2 cases are identical to the HEL1 cases except that they differ in their pitch profile as $r_{\rm c}$ is set to 0.5 for the HEL2 cases as opposed to $r_{\rm c} = 2.0$ for the HEL1 cases. The HEL2-B and C cases are for comparison with the reference case UNI-B whereas the HEL2-A case is for validation. The axial sonic Mach numbers corresponding to the HEL2-A, B, and C cases are $M_{\rm s}$ = 1.26, 5.0, and 10.0 respectively. 
    
\end{itemize}

The values of the parameters $B_{\rm 1} $ and $B_{\rm 0}$ are 1 and 0.25 respectively and are kept the same for all models with helical magnetic fields (see Eq.~\ref{eq:Equation9}). With the choice of physical scales used in the present work, the value of $B_{\rm z}$ at the axis corresponds to 44 $\mu$\,G and the value of the azimuthal component of the magnetic field is $B_{\rm \phi} = 71 \mu\,G$ at $r= R_{\rm j}$ for the HEL1 and HEL2 cases. We define the parameter $\beta = 2P/B^{2}$ to characterize the magnetic field strength for the simulation runs. For all the runs, we initially set $\beta=32$ corresponding to the values of gas pressure ($P_0$) and total magnetic field ($B_0$) on the jet axis. This initial value of $\beta$ defined at the axis will be the same for UNI cases within the plasma column, however, HEL1 and HEL2 cases will have a radial dependence of $\beta$. The initial value of $\beta$ at the jet radius $R_{\rm j}$ is $\beta(r = R_{\rm j}) = 7.37$ for the HEL1 cases and $\beta(r = R_{\rm j}) = 4.67$ for the HEL2 cases.
The UNI cases are subject to the purely hydro-dynamical KHI at the jet boundary while the HEL2 cases have a helical magnetic field which can hinder the KH modes on the jet surface.

Time is in units of radial Alfv$\Grave{e}$n crossing time which is defined as $t_{\rm a} = R_{\rm j}\sqrt\rho_{\rm 0}/B_{\rm 0}$. In the cases with jet density equal to that of the ambient ($\eta = 1$), the radial Alfv$\Grave{e}$n crossing time $t_{\rm a}$ corresponds to $\approx 78.3$\,kyr for the chosen set of physical scales. 
We ran the simulation for $A$ cases up-to 3.75 $t_{\rm a}$, while, the jet models UNI-B, HEL1-B, and HEL2-B were ran up to 4.75 $t_{\rm a}$ as the instabilities develop slightly later for high sonic Mach number flows as the growth rate of KHI depends inversely on the sonic Mach number \citep{Hardee2008}. The UNI-C and UNI-D cases have the same sonic Mach number ($M_{\rm s}$ = 5.0) as the B cases but higher jet-speed on-axis due to the low density $\rho_{\rm 0}$ that results in higher sound speed $c_{s}$. Consequently, these two cases were ran up to 7.25 $t_{\rm a}$ and 15.0 $t_{\rm a}$ as it takes a long time for the instabilities to develop due to higher jet speeds on-axis.

\section{Emission modelling}\label{emission}

Synchrotron and inverse-Compton (IC) are the two most primary radiation mechanisms responsible for the observed double humped spectral energy distribution (SED) in the jets.
To study the effects of magneto-hydrodynamic instabilities on the continuum emission spectra of jets at large scales, we have considered two different approaches to model the non-thermal emission. For the first approach, we have developed a post-processing tool that considers static particle spectra i.e., the spectra do not evolve with time.
For the second approach, we have used the hybrid macro-particle based framework in the {\sc{PLUTO}} code developed by \cite{Vaidya_2018}, where the particle spectra evolve with time depending on local fluid quantities, and the history is also preserved during its update.
In the following subsection, we have described the methods involved to calculate the Synchrotron (section \ref{Synchrotron}) and IC (section \ref{IC}) emissivity for the first approach with the static particle spectra along with the associated intensity and flux.

\subsection{Static particle spectra}\label{fixed_spectra}
In this approach, each grid cell of the computational domain is treated as a single emitting blob for which the emissivity is modeled. This considers a non-uniform distribution of mass density, magnetic fields and assumes a homogeneous energy distribution of the emitting ultra-relativistic particles for the calculation of grid distribution of emissivities. The inputs are the fluid variables obtained from the simulations done by using the {\sc{PLUTO}} code and the viewing angle. Additionally, the parameters that prescribe the static particle spectrum are also provided as inputs. For static power-law spectra (Eq.~\ref{eq:Equation23}), input parameters are the lower and upper energy bounds - $\gamma_{\rm min}$ and $\gamma_{\rm max}$ respectively, the power-law index $p$ and the ratio of the number density of the non-thermal electrons to fluid number density $\eta^{\rm NT}$ = $n_{\rm e}^{\rm NT} m_{\rm p}/ \rho$, where $n_{\rm e}^{\rm NT}$ is the number density of non-thermal electrons. 
The outputs are the intensity maps and flux (see Eq.~\ref{eq:Equation24} and Eq.~\ref{eq:Equation25}). The validation of this approach is given in the appendix \ref{app:validation}.

\subsubsection{Synchrotron emission}\label{Synchrotron}
The synchrotron emission from the jets may have either leptonic or hadronic or lepto-hadronic origin. In our work, we assume that the synchrotron emitting particles are mainly electrons and their energy distribution is a power-law with spectral index p.

Given the synchrotron power radiated by a single electron $P_{\rm syn}(\nu,\gamma)$, the total synchrotron emissivity due to an ensemble of ultra-relativistic electrons is computed using the following expression \citep{Longair2011}
\begin{equation}
    J_{\rm syn}(\nu,\hat {\bvec{n}}_{\rm los}) = \int P_{\rm syn}(\nu,\gamma)N(\gamma)d\gamma,
\label{eq:Equation16}
\end{equation}
where $N(\gamma)d\gamma$ is the total number of electrons per unit volume having a Lorentz factor in the range $\gamma$ to $\gamma + d\gamma$ with power-law spectral index p and is given by
\begin{equation}
    N(\gamma)d\gamma =  N_{\rm 0}\gamma^{\rm -p} d\gamma ,\hspace{10mm} \gamma_{\rm min} \hspace{1.5mm} < \hspace{1.5mm} \gamma \hspace{1.5mm} < \hspace{1.5mm} \gamma_{\rm max} 
\label{eq:Equation23}
\end{equation}
where $\gamma_{\rm min}$ and $\gamma_{\rm max}$ are the limits of the electron energies and $N_{\rm 0}$ is the normalization constant. \\

We used the fundamental radiative transfer equation in the optically thin limit \citep{Rybiki1986} to calculate the specific intensity. For this, the emissivity $ J_{\rm syn}(\nu, \hat {\bvec{n}}_{\rm los})$ (see \ref{eq:Equation17}) obtained at each grid cell (i.e., X$^{\prime}$, Y$^{\prime}$ and Z$^{\prime}$) is integrated for a given line of sight along the direction $\hat {\bvec{n}}_{\rm los}$ defined by the spherical coordinates $\theta$ and $\phi$ (i.e. $\hat {\bvec{n}}_{\rm los} = [\sin\theta\cos\phi, \sin\theta\sin\phi, \cos\theta]$) using
\begin{equation}
    I_{\rm \nu}(\nu, X^{\prime}, Y^{\prime}) = \int_{-\infty}^{\infty} J_{\rm syn}(\nu,X^{\prime},Y^{\prime},Z^{\prime})\, dZ^{\prime},
\label{eq:Equation24}
\end{equation} 
where we choose a Cartesian coordinate system with the $Z^{\prime}$-axis along the line of sight of the observer while the other two axes are in the sky plane.

Then the flux density at a particular frequency $\nu$ can be estimated by integrating the specific intensity distribution over the solid angle subtended at the observer's position by the projection of the emitting region in the jet on the sky plane. This is given by
\begin{equation}
    F_{\rm \nu}(\nu) = \int I_{\rm \nu}(\nu, X^{\prime}, Y^{\prime})\, d \Omega,
\label{eq:Equation25}
\end{equation}
where the solid angle is given by $d\Omega =(dX^{\prime} \times dY^{\prime})/D^{2}$. 

The total integrated flux density $F_{\rm \nu}(\nu)$ can then be used to plot the continuum emission spectra. Flux density is scaled with ${F_{\nu_{\rm sc}}}$,
\begin{equation}
    {F_{\nu_{\rm sc}}} = \frac{E_{\rm sc}c}{{r_{\rm L}}^3\nu_{\rm sc}} = 4\pi I_{\nu_{\rm sc}},
\end{equation} where, $E_{\rm sc} = \gamma_{\rm sc}m_{e}c^{2}$ is the energy scale where $\gamma_{\rm sc} = 1$, and frequency is scaled in units of Larmor frequency $\nu_{sc} = \nu_{\rm G} \approx 122$Hz which is estimated using the initial magnetic field strength defined at the axis of the plasma column (i.e., $B_z(t=0)$). Further, $r_L$ is the Larmor radius for highly relativistic electrons, with Lorentz factor $\gamma_{\rm max}$, and is obtained using the same initial axial field strength.
For the chosen set of physical scales, the value of the flux scale in physical units is ${F_{\nu_{\rm sc}}} \approx 9.38\times10^{-8} \left(\frac{\gamma_{\rm max}}{10^{6}}\right)^{-3}$ ergs s$^{-1}$ cm$^{-2}$ Hz$^{-1}$.
In this work, we estimate the synthetic emission assuming a reference distance $D = 7.9$ Mpc between the source and the  observer. As a result, considering the length scale ($R_{\rm j}$) and the grid resolution we obtain the solid angle as
\begin{equation}
    \frac{d\Omega}{4\pi} = 5.1 \times 10^{-15} \left(\frac{D}{7.9 \rm Mpc}\right)^{-2}
\end{equation}

\subsubsection{IC emission}\label{IC}
Inverse-Compton scattering involves two types: synchrotron self-Compton (SSC) and external Compton (EC). As the focus is on studying the portion of the jet far away from the central black hole, the seed photons for the inverse-Compton scattering are typically from the Cosmic Microwave Background (CMB). 
We assume an isotropic seed photon distribution and the bulk-flow of the jet at kpc-scales to be in the non-relativistic limit. The total IC emissivity at a particular frequency due to this ensemble of ultra-relativistic electrons can be computed using the form given in \cite{Petruk2009}
\begin{equation}
    J_{\rm IC}(\nu) = \int P_{\rm IC}(\nu,\gamma)N(\gamma)d\gamma,
\label{eq:Equation19}
\end{equation}
where, $P_{\rm IC}(\nu, \gamma)$ is the IC power radiated due to a single electron. 
The IC intensity maps and flux density are calculated by using the same methods as explained in section \ref{Synchrotron} using the expression of emissivity from equation \ref{eq:Equation20} at each grid cell in the numerical domain.

\subsection{Evolving particle spectra}

One of the limitations of our first approach which makes it a rather simplistic prescription for emission modeling is the assumption of a static power-law distribution of relativistic electrons (i.e., a constant value of power-law index $p$). 
In order to relax the above constraint and take into account the effects of micro-physical processes (for example particle acceleration due to shocks) on the distribution function of emitting particles and subsequently on emissivities that in general depends on the grid position:  
$J_{\rm syn}(\nu,\hat {\bvec{n}}_{\rm los}$, X$^{\prime}$, Y$^{\prime}$, Z$^{\prime}$) and  $J_{\rm IC}(\nu,\hat {\bvec{n}}_{\rm los}$, X$^{\prime}$, Y$^{\prime}$, Z$^{\prime}$), we use another approach where the energy spectra of the emitting particles evolve with time. 

It follows a Lagrangian macro-particle based approach where each of these macro-particles is essentially an ensemble of non-thermal particles (e.g. electrons in this case). The outputs of the hybrid model are the Synchrotron and IC emissivities that are generated by evolving the particle spectra for a user-defined observing frequency value. 
These emissivities are provided as inputs to the post-processing tool to obtain the intensity maps and flux using equations \ref{eq:Equation24} and \ref{eq:Equation25}. The detailed methodology of this approach and the equations considered to calculate the emissivities are given in \cite{Vaidya_2018}. We initialize all the runs listed in Table~\ref{tab:Table1} using the hybrid framework in the {\sc{PLUTO}} code with $6\times 10^{5}$ Lagrangian macro-particles that are randomly distributed in space following Gaussian deviates that depend on the cylindrical radius. 
This allows complete sampling of the plasma column as more particles are initialized close to the axis. 
The initial electron distribution is chosen as a power-law given by Eq.~\ref{eq:Equation23} in which the normalization constant is
\begin{equation}
    N_0 = \frac{\eta^{\rm NT} \rho (p - 1)}{m_{\rm p} \left(\gamma_{\rm min}^{\rm 1-p} - \gamma_{\rm max}^{\rm 1-p}\right)}.
\end{equation}

The electron distribution within each macro-particle is evolved depending on the physical grid quantities interpolated at the macro-particle position. In particular, for each macro-particle, the Fokker-Planck equation is solved accounting for the radiative losses due to adiabatic expansion, synchrotron, and IC-CMB emission. At kpc scales in jets, stochastic acceleration of particles due to turbulence may also contribute to the evolution of the particle spectra and the diffuse emission at high energies \citep{Hess_2020}. However, for the present work, we have not accounted for the diffusion of particles in momentum space (Fermi IInd order process). 
Additionally, the spatial diffusion of electrons is also neglected.
On neglecting the above diffusion terms, the adopted approach ensures that the total number of electrons remains constant within a single macro-particle.

To account for particle acceleration at shocks, the spectral distribution of any macro-particle experiencing the shock is updated based on the compression ratio of the shock. To estimate the compression ratio, the shocked regions are flagged using conditions on the pressure gradient threshold and negative value of velocity divergence. Further, the macro-particles entering into such shocked regions quantify the pre-shock and post-shock conditions from interpolated fluid values. These conditions are then used to compute the orientation of the shock normal and thereafter the shock speed using the co-planarity theorem \citep{Schwartz_1998}. Finally, the compression ratio $s$ is estimated as the ratio of upstream and downstream velocities in the shock rest frame.  
It should be noted that in this work, only shocks are considered as a possible re-acceleration mechanism and no distinction is made with regards to quasi-parallel or quasi-perpendicular shocks. As the shocks encountered are non-relativistic, the standard diffusive shock acceleration (DSA) approach that assumes isotropic particle distribution has been adopted whereby the spectral slope flattens to a momentum index given as $q= 3s/(s-1)$ \citep[e.g.,][]{Blandford_1978, Drury_1983, Kirk_2000, Achterberg_2001}.
Additionally, the maximum energy of the updated spectra is estimated by equating the acceleration time scale with the radiative cooling time scale that depends on the strength of the magnetic field $B$. It can be expressed as follows:
\begin{equation}
  \gamma_{\rm max} = \left(\frac{9c^{4}m_{\rm e}^{2}}{8 \pi B e^{3}}\right)^{1/2}.
\end{equation}

Using the maximum energy of the updated spectra and the compression ratio, the particle distribution is updated by following \citep[e.g.,][Mukherjee et. al. (in prep)]{Jones1994, Micono1999, Winner_2019}
\begin{equation}
    N_{\rm out}(\gamma) = q \int_{\gamma_{\rm min}}^{\gamma} N_{\rm in}(\gamma^\prime) \left(\frac{\gamma}{\gamma^{\prime}}\right)^{-q+2} \frac{d\gamma^{\prime}}{\gamma^{\prime}},
\end{equation}
where, $\gamma \in [\gamma_{\rm min}, \gamma_{\rm max}]$, and $N_{\rm out}$ is the updated spectra in the post-shock region which is dependent on the incoming (pre-shock) particle spectra $N_{\rm in}$.
Such an update of the particle spectrum allows preserving the history of the particle distribution as it traverses the shock. 

At any given time, the instantaneous distribution of electrons for each macro-particle is convolved with single electron power for estimating emissivity associated with that macro-particle 
Finally, the value of emissivity for each macro-particle is interpolated back onto the grid. Once we can obtain the grid distribution of the emissivities $J_{\rm syn}(\nu,\hat {\bvec{n}}_{\rm los}$, X$^{\prime}$, Y$^{\prime}$, Z$^{\prime}$) and  $J_{\rm IC}(\nu,\hat {\bvec{n}}_{\rm los}$, X$^{\prime}$, Y$^{\prime}$, Z$^{\prime}$), we integrate them along a line of sight to generate the intensity maps and emission spectra using the same method described in section~\ref{Synchrotron}.

\section{Results}\label{Results}

The methods of dynamical and emission modeling of magneto-hydrodynamic instabilities in jets have been described in sections \ref{setup} and \ref{emission} respectively. 
This section comprises the results obtained from the dynamical modeling and emission modeling and gives a detailed discussion of our major findings of this work. Sections \ref{sec:emission_NoShocks}, \ref{sec:emission_WithShocks} $\&$ \ref{sec:emission_Comp} elaborate on the effects of the MHD instabilities on the observed jet emission based on different scenarios.

\subsection{Dynamical modeling results}\label{dynamics_results}

\begin{figure*}
\centering
\subfigure{\includegraphics[width=6.5cm]{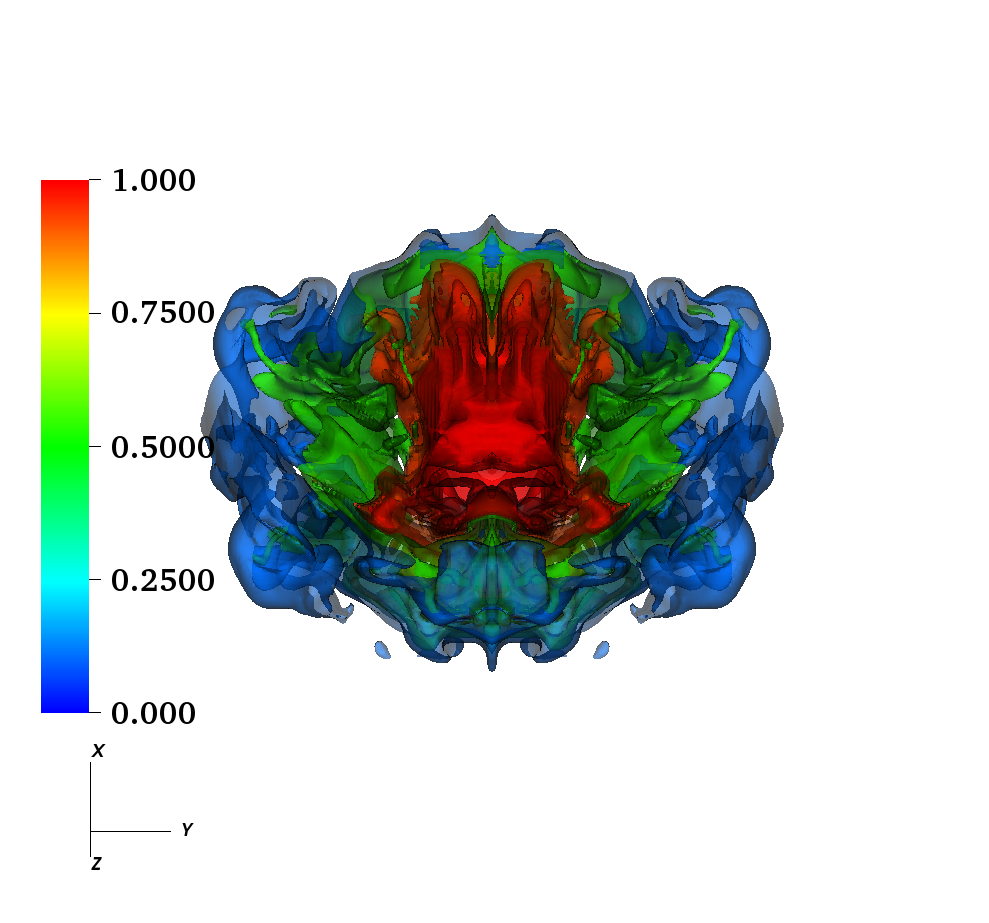}}
\subfigure{\includegraphics[width=6.5cm]{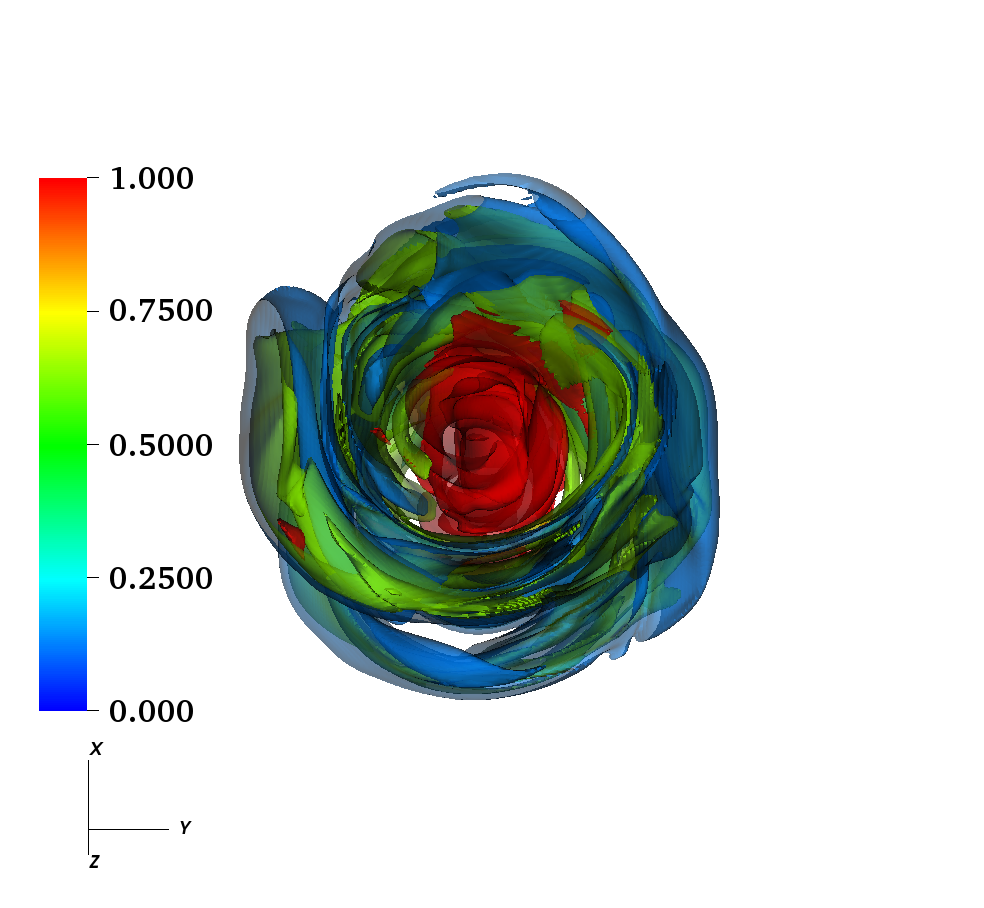}}
\subfigure{\includegraphics[width=6.5cm]{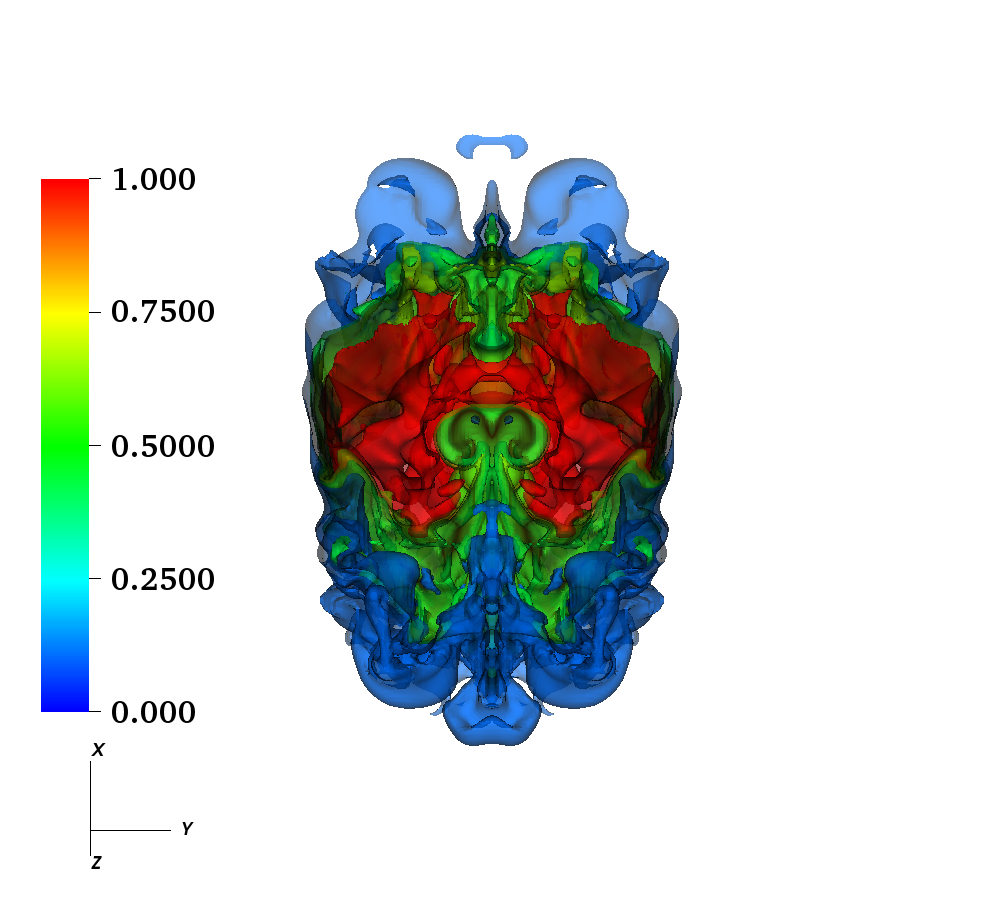}}
\subfigure{\includegraphics[width=6.5cm]{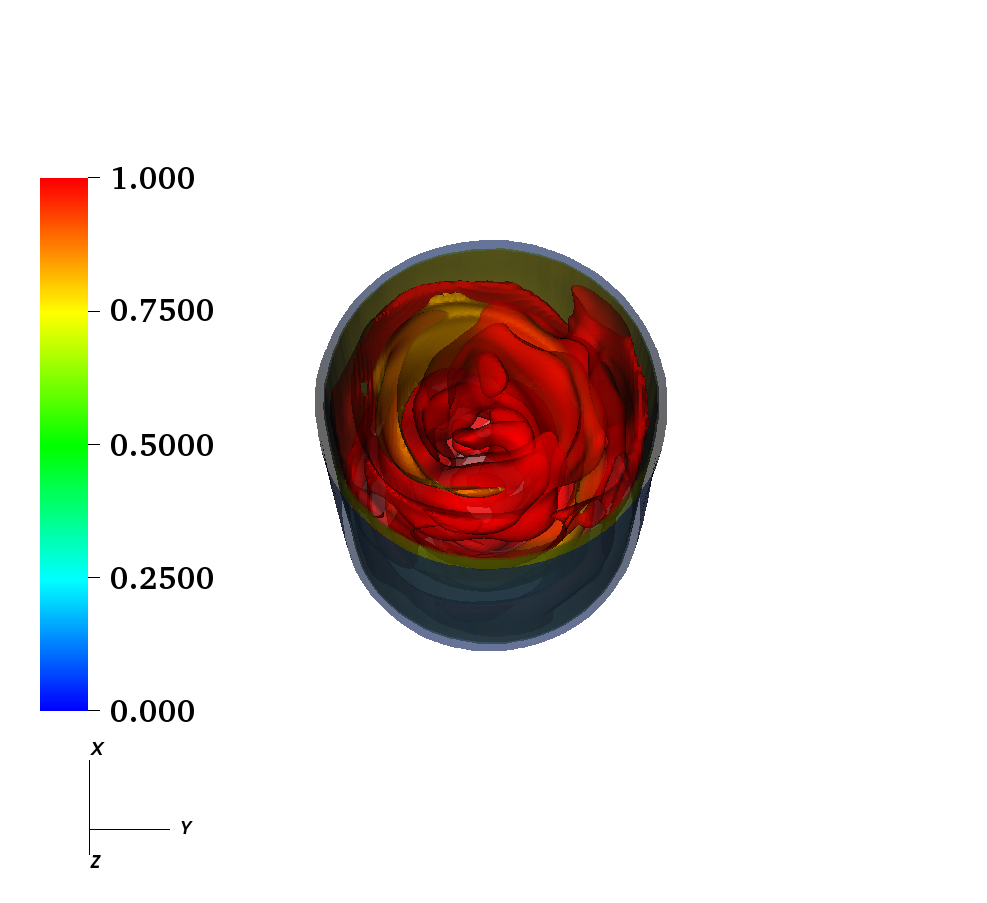}}
\caption{The three-dimensional isosurface density contours of the model UNI-A (top-left), HEL2-A (top-right), UNI-B (Bottom-left,) and HEL2-B (Bottom-right) at $t/t_{\rm a} = 3.75$ for a direction along a line of sight inclined at 20$^{\circ}$ with the jet axis. The colorbars represent the jet density $\rho/\rho_{\rm 0}$. While the UNI cases have a turbulent flow structure as they experience the KH instability alone, the HEL2 cases have a clearly identifiable jet boundary as they are relatively stable.  
}
\label{fig:Figure2}
\end{figure*}

We do the dynamical modeling of jets to study the effects of magneto-hydrodynamic instabilities on the jet dynamics and energetics. By solving the ideal MHD equations, we study the temporal evolution of three-dimensional scalar and vector fields such as mass density $\rho$, gas pressure $P$, magnetic field $\bvec{B}$, and velocity field $\bvec{v}$. Three-dimensional snapshots of the jet density structure for the models UNI-A, UNI-B, HEL2-A, and HEL2-B at $t/t_{\rm a} = 3.75$ are shown in Fig.~\ref{fig:Figure2} for a direction lying in the X-Z plane (i.e. $\hat {\bvec{n}}_{\rm los} = [\sin\theta, 0, \cos\theta]$) along a line of sight inclined at $\theta = 20^{\circ}$ with the jet axis.
In the uniform magnetic field configurations i.e., UNI-A and UNI-B where only the Kelvin-Helmholtz instability is present, the jet dynamics are greatly altered by $t/t_{\rm a} = 3.75$ as compared to the helical configurations HEL2-A and HEL2-B in which the jet remains relatively stable as the jet boundary is clearly distinguishable.

\begin{figure*}
\centering
\subfigure{\includegraphics[width=6.5cm]{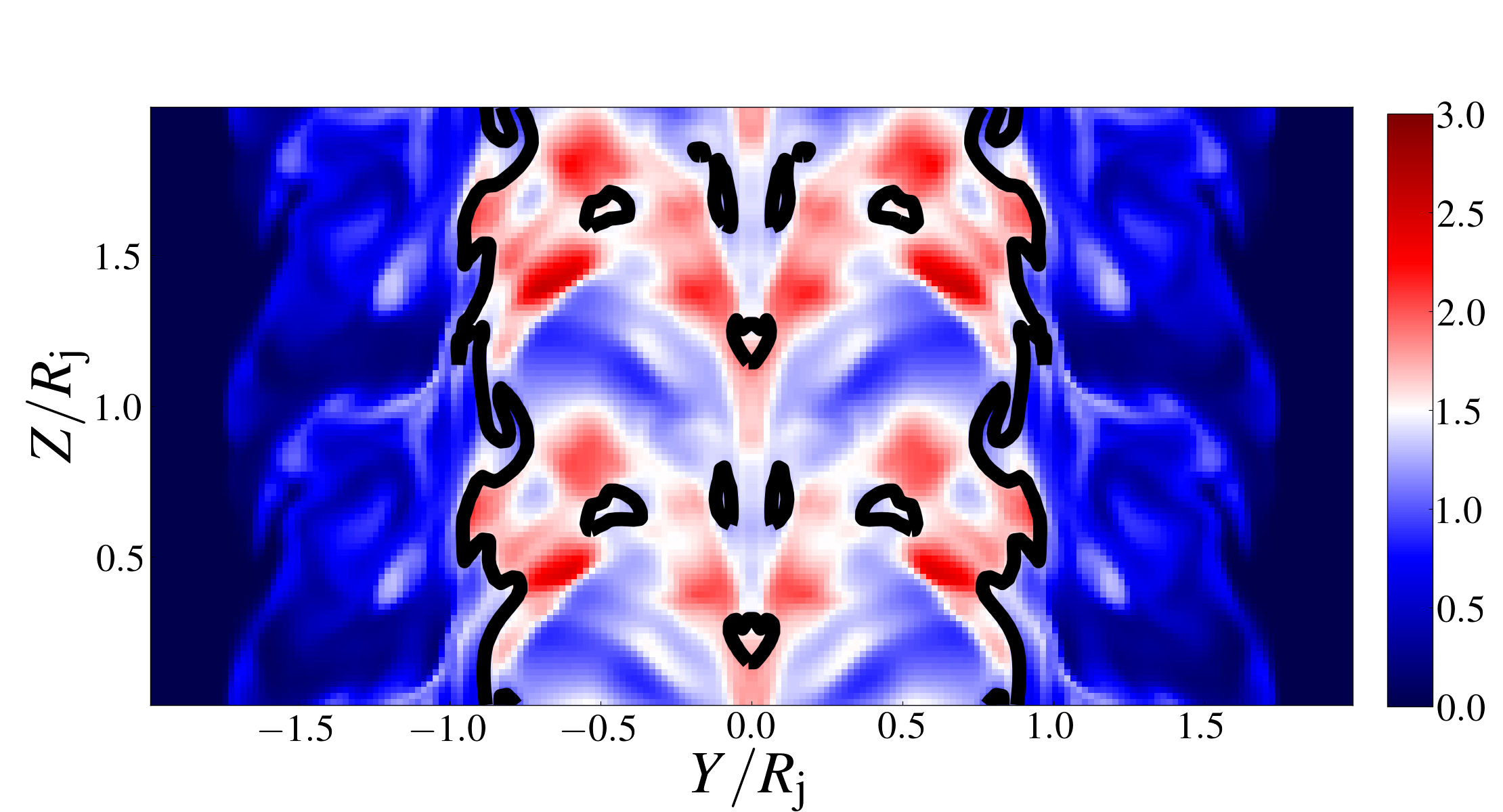}}
\hskip3ex
\subfigure{\includegraphics[width=6.5cm]{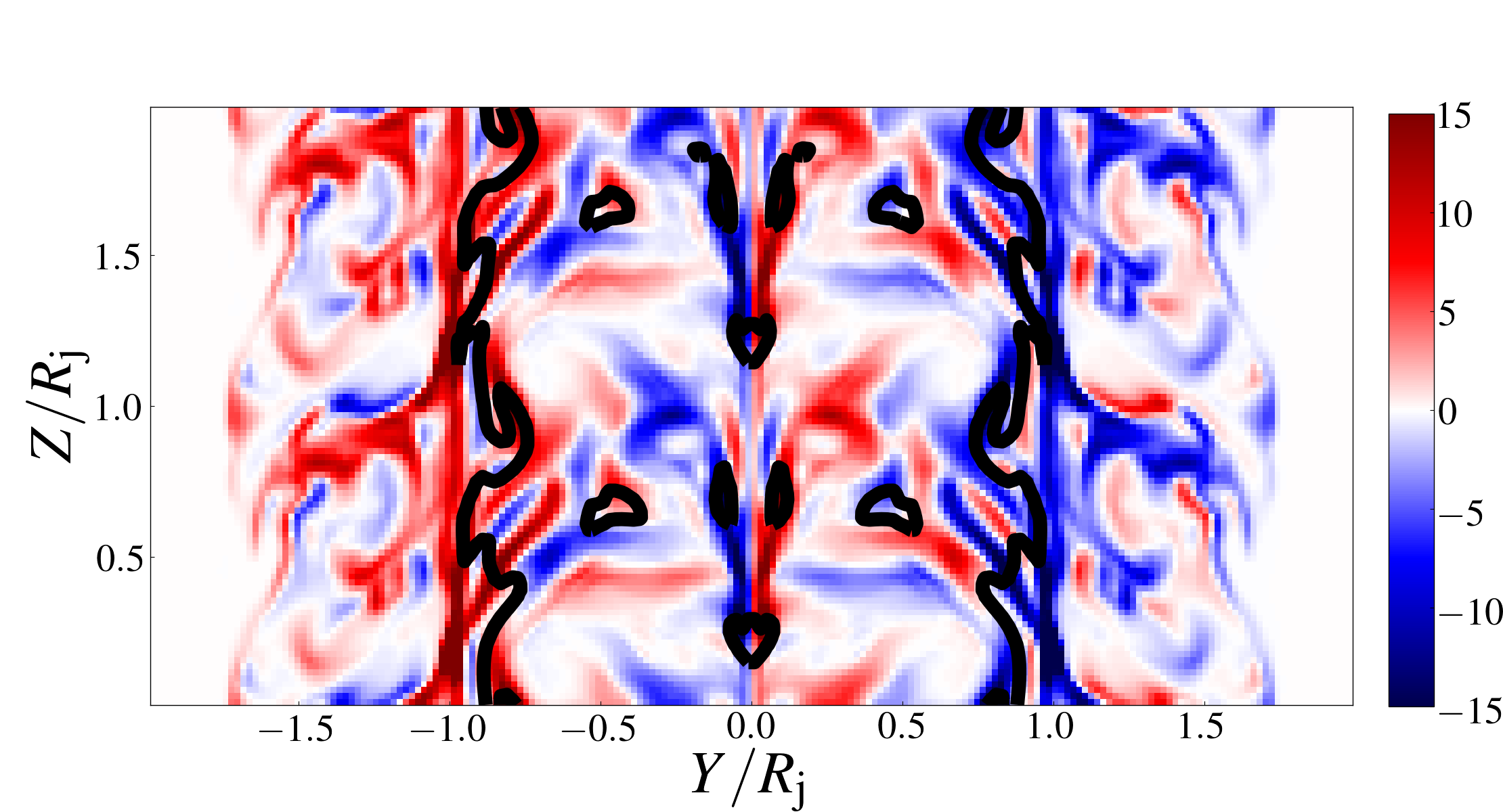}}
\subfigure{\includegraphics[width=6.5cm]{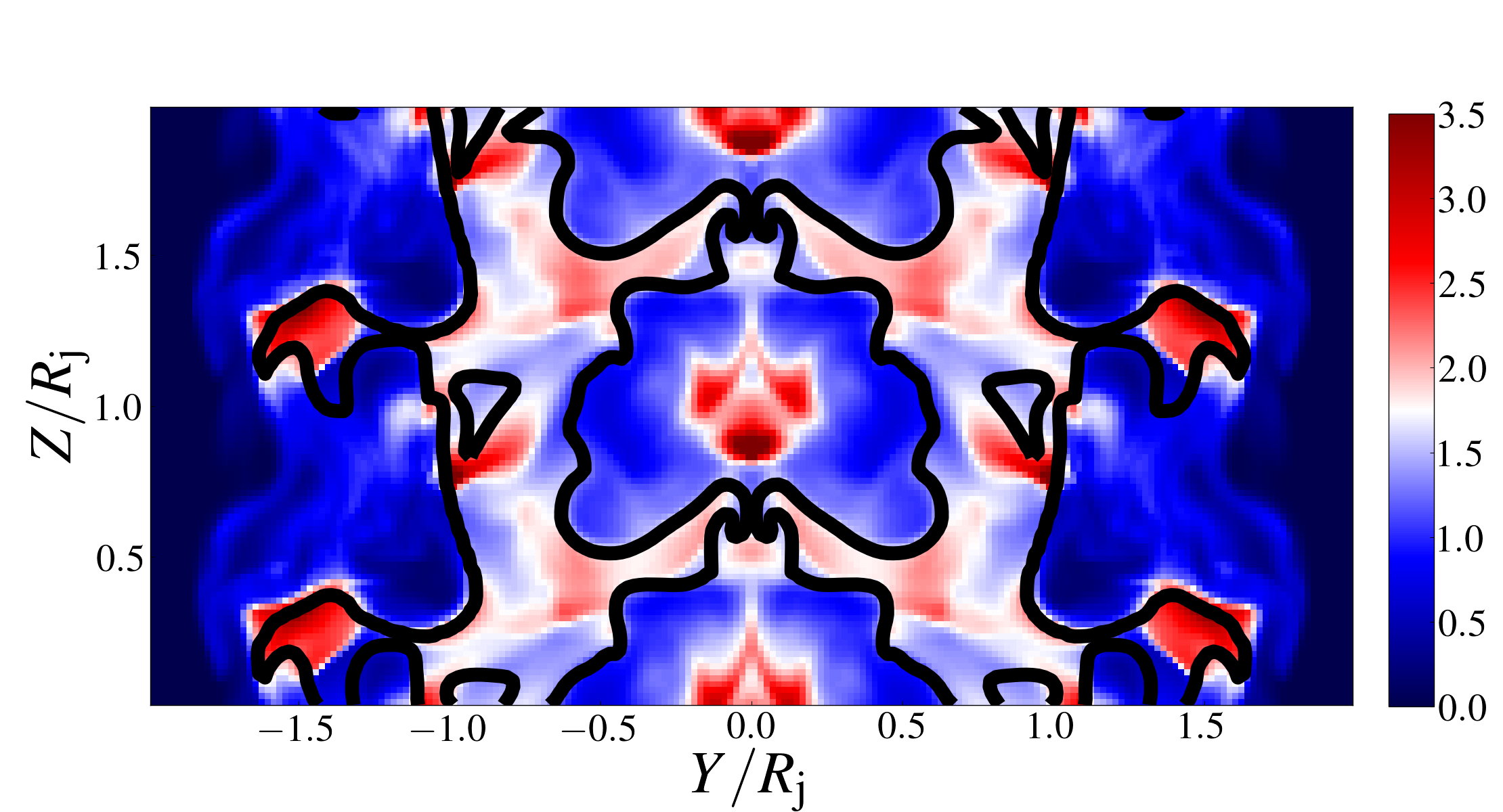}}
\hskip3ex
\subfigure{\includegraphics[width=6.5cm]{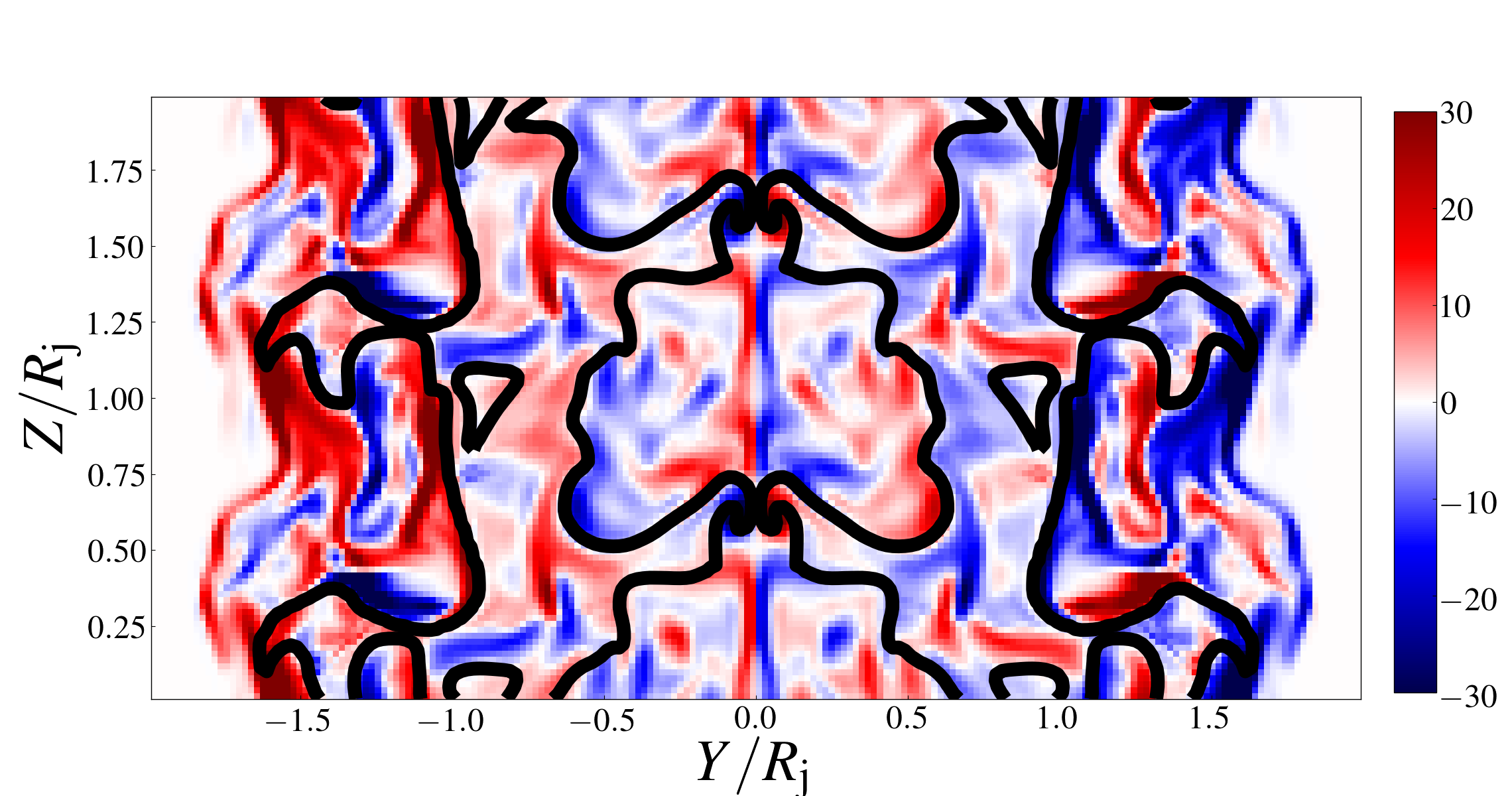}}
\subfigure{\includegraphics[width=6.5cm]{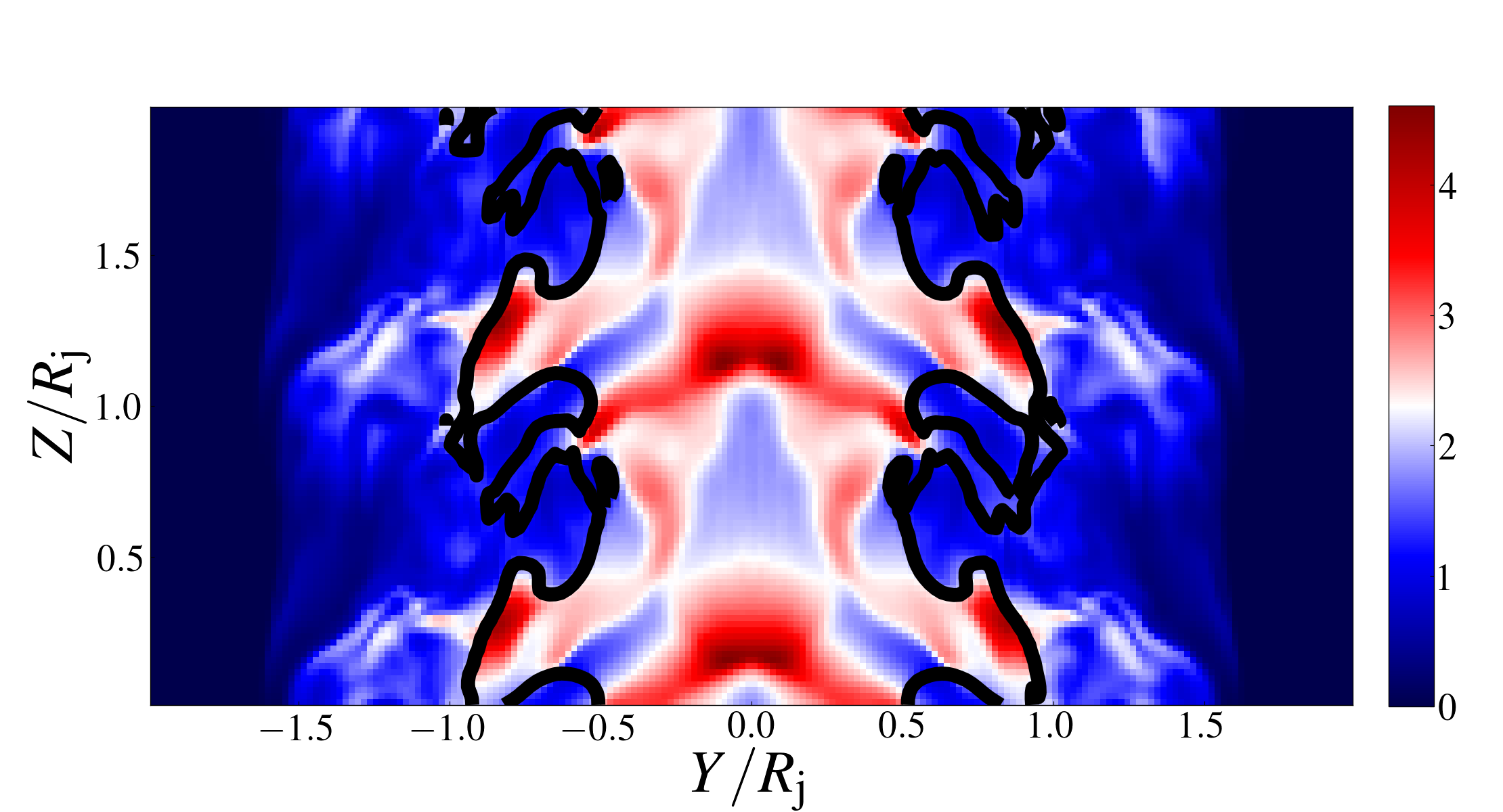}}
\hskip3ex
\subfigure{\includegraphics[width=6.5cm]{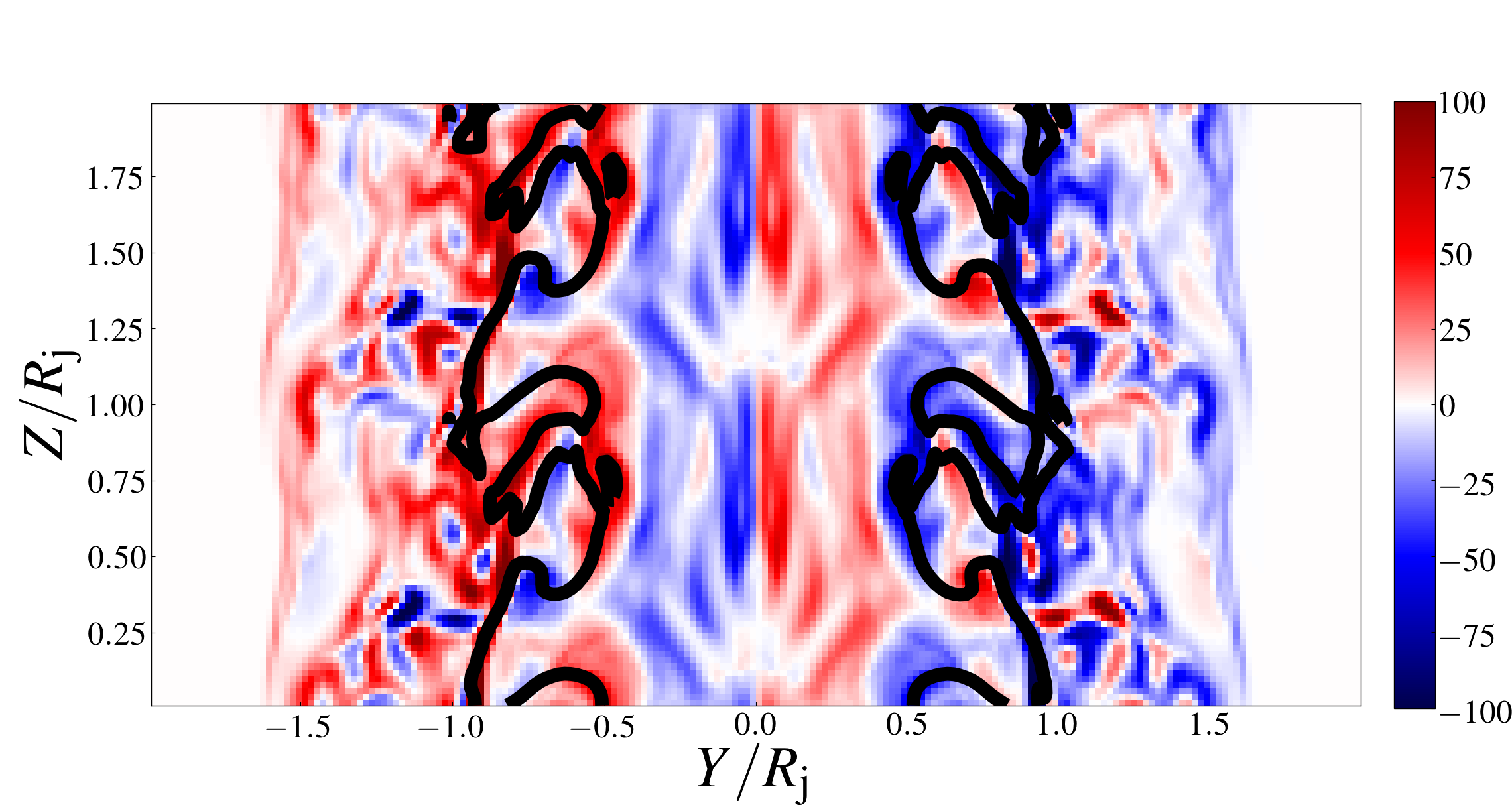}}

\caption{Left-hand panels: The distribution of jet pressure $P/P_{0}$ in the Y-Z plane of the models UNI-B at $t/t_{\rm a} = 3.75$ (top), UNI-C at $t/t_{\rm a} = 5.59$ (middle), and UNI-D at $t/t_{\rm a} = 10.61$ (bottom) respectively. Right-hand panels: The distribution of the vorticity along the normal to the Y-Z plane $(\nabla \times \bvec{v})_{\rm x}$ for the corresponding cases at the same times. The colorbars in the left and right panels represent the corresponding pressure and vorticity respectively in code units. Overplotted as black lines are the tracer contours at level 0.8 to demarcate the jet boundary in all three cases. The tracer levels close to the jet axis are greater than 0.8 and are not shown here.}
\label{fig:Figure3_new}
\end{figure*}

Fig.~\ref{fig:Figure3_new} shows the two-dimensional cuts of pressure and vorticity distribution in the Y-Z plane for the UNI-B case at $t/t_{\rm a} = 3.75$, for the UNI-C case at $t/t_{\rm a} = 5.59$, and for the UNI-D case at $t/t_{\rm a} = 10.61$ respectively. In the UNI-B case, the KH instability in the shear layer at the jet boundary results in the formation of strong shocks. Such shocked structures are also seen on the jet boundary in the UNI-C and UNI-D cases. Additionally, for these cases, there are regions at high pressure on the jet axis due to the formation of biconical shocks which are relatively weaker as compared to those formed at the jet boundary.  A detailed discussion on the under-dense jets UNI-C and UNI-D is presented in section \ref{under_dense_SED}. 

In all three cases, the vortical structures are present at the jet boundary as they can only form in regions with velocity shear. The vorticity is highest in the UNI-D case as there is a large velocity shear due to high axial velocity inside the jet ($V_{z} = 0.7c$).
The strong shocks resulting from the high-velocity shear influence the emission properties of MHD jets that are discussed in section  \ref{sec:emission_WithShocks}.

\subsubsection{Jet energetics and validation of dynamical jet models}

\begin{figure*}
\centering
\includegraphics[width = 16.2cm, scale=1.0]{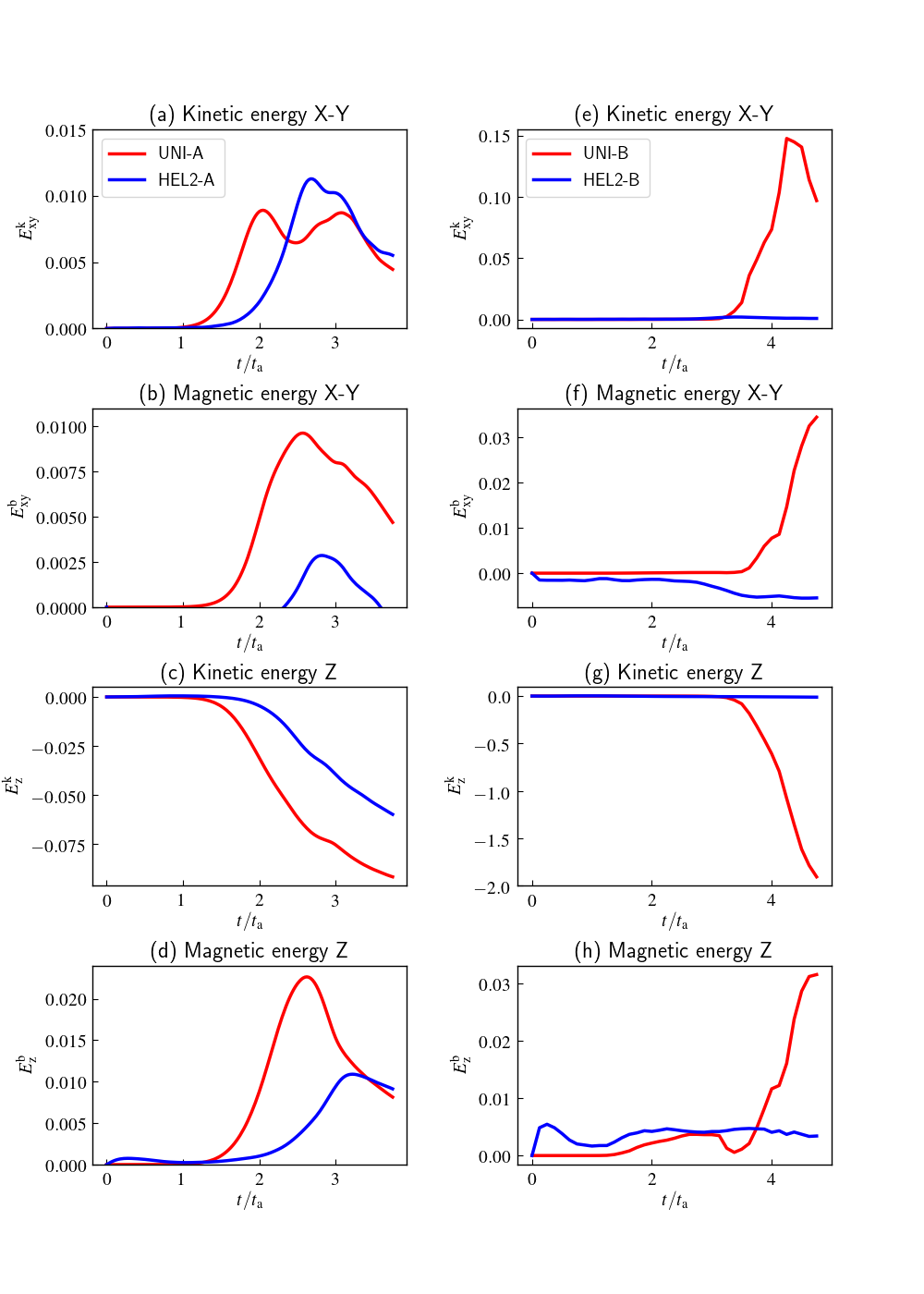}
\caption{Left-hand panels: The time evolution of the volume-averaged perturbed (a) kinetic energy in the X-Y plane $E^{\rm k}_{\rm xy}$, (b) magnetic energy in the X-Y plane $E^{\rm b}_{\rm xy}$, (c) axial kinetic energy $E^{\rm k}_{\rm z}$, and (d) axial magnetic energy $E^{\rm b}_{\rm z}$ for the UNI-A and HEL2-A cases. Right-hand panels: The time evolution of corresponding energies for their higher sonic Mach number ($M_{\rm s} = 5.00$) counterparts.}
\label{fig:Figure3}
\end{figure*}

In order to validate our dynamical jet models, we set up our $A$ cases with initial conditions identical to those given for the three configurations in \cite{Baty_2002}. For comparison, we plotted the time evolution of the volume-averaged energies defined below (see Fig.~\ref{fig:Figure3}).

The volume-averaged perturbed kinetic energy confined to the X-Y plane $E^{\rm k}_{\rm xy}$ is given by 
\begin{equation}
    E^{\rm k}_{\rm xy} = \frac{1}{V_{\rm b}} \int_{V_{\rm b}} \frac{\rho V_{\rm x}^{2} + \rho V_{\rm y}^{2}}{2} \hspace{1mm} \hspace{1mm}dX\hspace{1mm}dY\hspace{1mm}dZ - E^{\rm k,0}_{\rm xy},
\end{equation}
and the volume-averaged perturbed magnetic energy confined to the X-Y plane $E^{\rm b}_{\rm xy}$, is 
\begin{equation}
    E^{\rm b}_{\rm xy} = \frac{1}{V_{\rm b}} \int_{V_{\rm b}} \frac{B_{\rm x}^{2} + B_{\rm y}^{2}}{2} \hspace{1mm} \hspace{1mm}dX\hspace{1mm}dY\hspace{1mm}dZ - E^{\rm b,0}_{\rm xy},
\end{equation}
where $V_{\rm b} = 16 R_{\rm j}^{2} L_{\rm z}$ is the volume of the simulation box, and $E^{\rm k,0}_{\rm xy}$ and $E^{\rm b,0}_{\rm xy}$ are the initial kinetic and magnetic energies obtained from the equilibrium conditions. As $E^{\rm k,0}_{\rm xy} = 0$, the volume-averaged perturbed kinetic energy confined to the X-Y plane $E^{\rm k}_{\rm xy}$, is the same as volume-averaged kinetic energy. The volume-averaged perturbed axial kinetic and magnetic energies are given by  

\begin{equation}
    E^{\rm k}_{\rm z} = \frac{1}{V_{\rm b}} \int_{V_{\rm b}} \frac{\rho V_{\rm z}^{2}}{2} \hspace{1mm} \hspace{1mm}dX\hspace{1mm}dY\hspace{1mm}dZ - E^{\rm k,0}_{\rm z},
\end{equation}
and
\begin{equation}
    E^{\rm b}_{\rm z} = \frac{1}{V_{\rm b}} \int_{V_{\rm b}} \frac{B_{\rm z}^{2}}{2} \hspace{1mm} \hspace{1mm}dX\hspace{1mm}dY\hspace{1mm}dZ - E^{\rm b,0}_{\rm z}.
\end{equation}

We extend this analysis to the jet configurations with a higher sonic Mach number ($M_{\rm s} = 5$) (i.e., $B$ cases). The results for the UNI-A and HEL2-A cases ($M_{\rm s} = 1.26$) shown in the left-hand panels of Fig.~\ref{fig:Figure3} are in agreement with those obtained by \cite{Baty_2002} and the results for the corresponding B cases are shown in the right-hand panels of Fig.~\ref{fig:Figure3}. 
In the jet configuration UNI-A where only the Kelvin-Helmholtz instability is present, vortex growth occurs at the jet boundary \citep{Baty_2002}. The magnetic field lines get concentrated around the edges of these vortices and the build-up of magnetic and kinetic energy $E^{\rm b}_{\rm xy}$ and $E^{\rm k}_{\rm xy}$, in the transverse direction, disrupts the flow (see panels (a) and (b) in Fig.~\ref{fig:Figure3}). In contrast, the vortex growth is suppressed by the azimuthal magnetic field $B_{\rm \phi}$ which stabilizes the jet in the HEL2-A configuration. In the B cases with $M_{\rm s} = 5$, the beginning of the instability is marked by the deviation of the axial kinetic energy from its initial value which occurs at $t/t_{\rm a} \approx 3.375$ in the UNI-B case  
(see panel (g) in figure~\ref{fig:Figure3}). In the UNI-B case, the Kelvin-Helmholtz instability makes the flow turbulent at small scales resulting in freshly formed shocks that disrupt the flow. As a result, the kinetic and magnetic energies in the transverse direction, $E^{\rm k}_{\rm xy}$ and $E^{\rm b}_{\rm xy}$ increase from $t/t_{\rm a} \approx 3.375$ (see top two panels in the right column of figure~\ref{fig:Figure3}). 
In the HEL2-B case, small deviations are seen in the transverse and axial magnetic energies, $E^{\rm b}_{\rm xy}$ and $E^{\rm b}_{\rm z}$ as the presence of a helical magnetic field suppresses the steepening of any turbulent features thereby curbing any shock formation.

\subsection{Emission modeling: No shocks case} \label{sec:emission_NoShocks}

For emission modeling, we adopt two different mechanisms viz., static and evolving particle spectra. 
Modeling emission with evolving particle spectra helps in incorporating the additional physics due to radiative losses.

\begin{figure*}
\centering
\includegraphics[width = 18.0cm, scale=1.0]{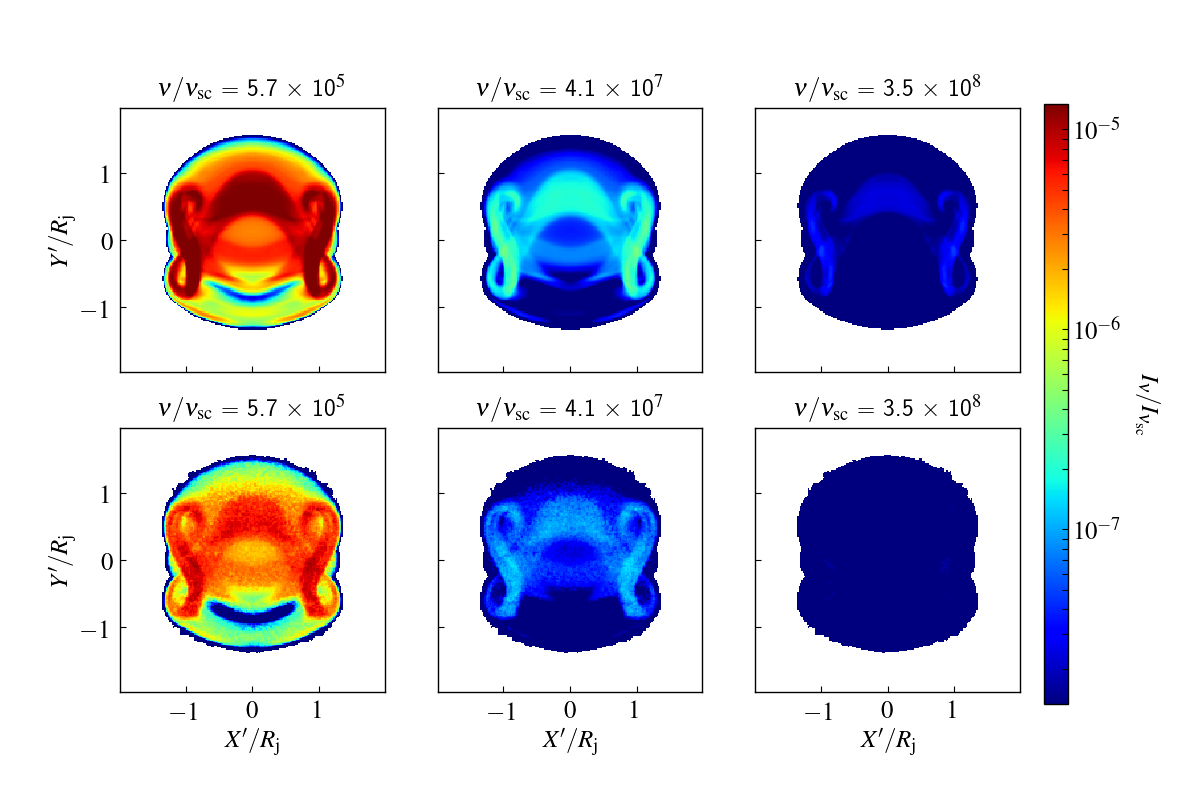}
\caption{
Synchrotron emission maps produced using the static particle spectra (top panels) and evolving particle spectra (bottom panels) with an initial power-law index, $p=3$ for case UNI-A at $t/t_{\rm a}$ = 2.25  projected on the sky plane $X^{\prime}$-$Y^{\prime}$. These maps are obtained for a direction along a line of sight inclined at 20$^{\circ}$ with the jet axis at three different observing frequencies normalized to the frequency scale $\nu_{\rm sc} = 122$ Hz.
The colorbar represents the magnitude of specific intensity $I_{\rm \nu}$ normalized to the specific intensity scale $I_{\rm \nu_{\rm sc}} = 7.46 \times 10^{-9}$ ergs s$^{-1}$ cm$^{-2}$ Hz$^{-1}$ str$^{-1}$}.
\label{fig:Figure7_new}
\end{figure*}

Fig.~\ref{fig:Figure7_new} shows the synchrotron emission maps for the UNI-A case obtained at $t/t_{\rm a} = 2.25$ using the static particle spectra (top panels of Fig.~\ref{fig:Figure7_new}) and evolving particle spectra (bottom panels of Fig.~\ref{fig:Figure7_new}). 
In the A cases, for both approaches, we assume that the energy distribution of the ultra-relativistic emitting electrons is a power-law with spectral index $p=3$, and energy limits $\gamma_{\rm min} = 100$  and $\gamma_{\rm max} = 10^{6}$. The ratio of the number density of the non-thermal electrons to fluid number density is taken as $\eta^{\rm NT} = 10^{-3}$.
As the initial distribution of non-thermal particles is a power-law for both approaches, the number of emitting electrons drops with an increase in energy which leads to a dimming effect in the intensity maps.
The two bright features resembling a figure of eight which appear in these emission maps can be attributed to the magnetic field structure at $t/t_{\rm a} = 2.25$. 
The Kelvin-Helmholtz instability in the UNI-A case leads to vortex formation at the jet boundary. The magnetic field lines get concentrated at the edges of the vortices due to a local increase in jet density which results in increased emission from these regions.
An enhanced dimming is seen in the intensity maps obtained with evolving particle spectra (bottom panels of Fig.~\ref{fig:Figure7_new}) at higher frequencies as the bright features completely vanish in the emission map obtained at $\nu/\nu_{\rm sc} = 3.5 \times 10^8$. This difference is purely on the account of radiative losses due to the synchrotron process which is unaccounted for in the static particle spectra approach. 
Further, the UNI-A case which has a relatively lower initial axial sonic Mach number does not show any shock feature to energize particles undergoing radiative losses. 
Therefore, the non-thermal electrons lose energy with time due to the synchrotron cooling effect which results in the enhanced dimming of the jet emission at higher energies as cooling becomes more efficient at higher energies due to shorter cooling time. Using the evolving particle spectra, no shocks were captured in any of the A cases. For studying the impact of shocks, we ramp up the initial sonic Mach number to $M_{\rm s} = 5$ along the jet axis for all three configurations to study its effects on the jet emission. 

\subsection{Emission modeling: Cases with shocks} \label{sec:emission_WithShocks}
Initially, the reference case UNI-B with axial sonic Mach Number $M_{\rm s}$ = 5, has uniform density, pressure, and axial magnetic field. The growth of the KH instability leads to a highly turbulent flow structure as shown in Fig.~\ref{fig:Figure2}. Vortex formation occurs near the shear layer due to the turbulence (see top-right panel in figure~\ref{fig:Figure3_new}). This leads to the formation of shocks near the shear layer.

\begin{figure*}
\centering
\subfigure{\includegraphics[width = 13.5 cm]{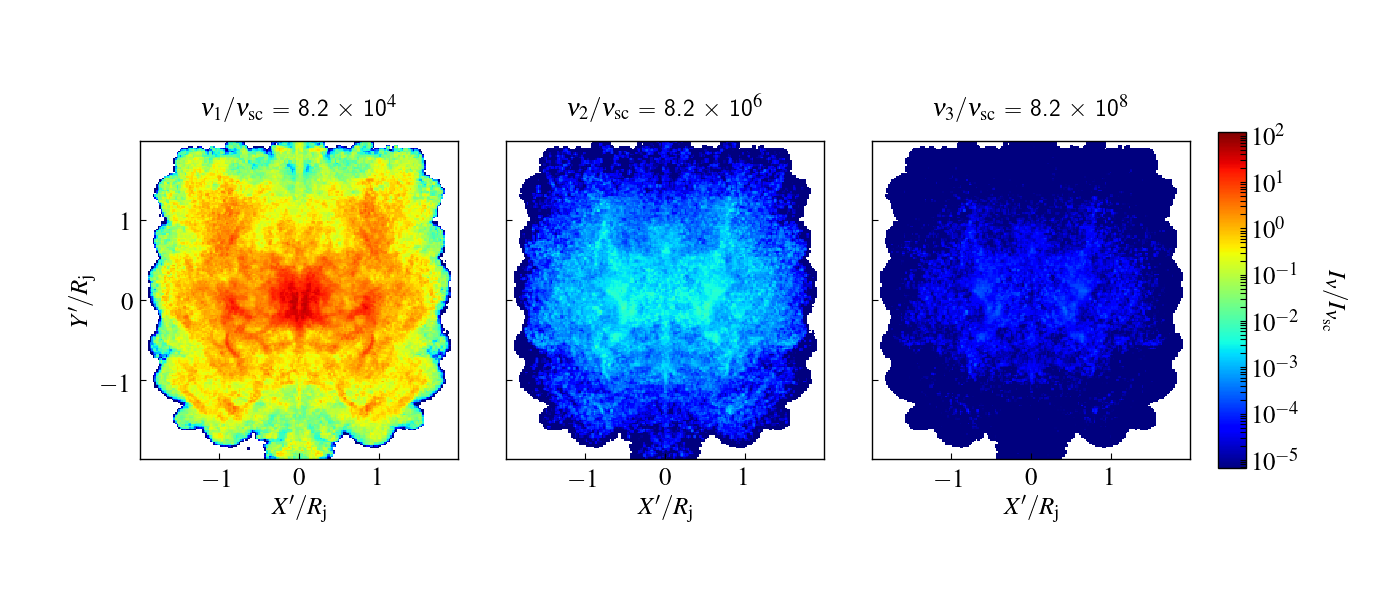}}
\vskip-7ex
\label{fig:Figure7A}
\subfigure{\includegraphics[width = 13.5 cm]{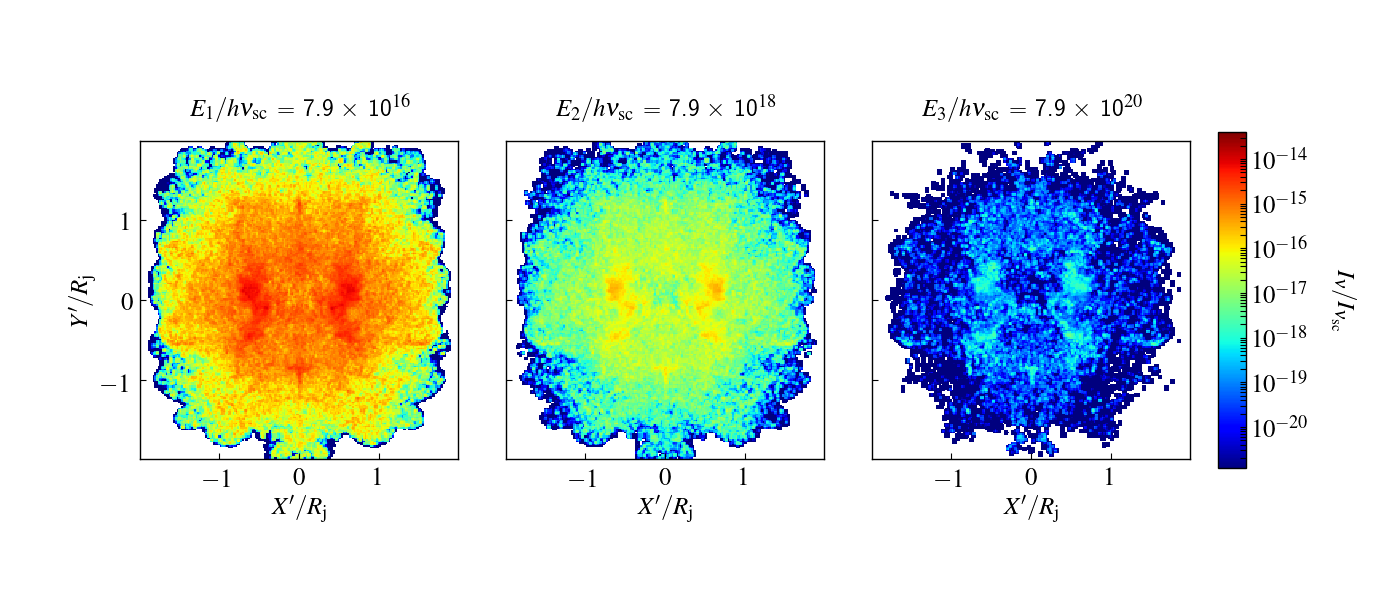}}
\vskip1ex
\label{fig:Figure7B}
\hskip1ex
\subfigure{\includegraphics[width = 13.5 cm]{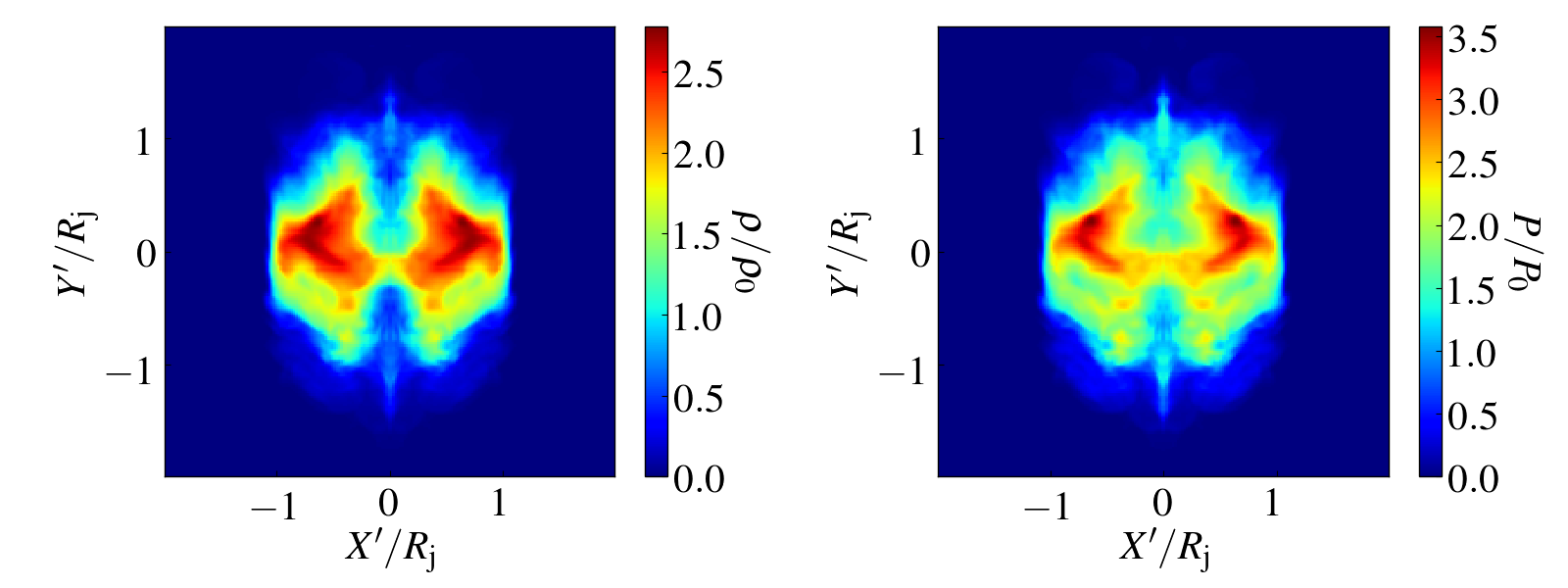}}
\caption{Top panels: Synchrotron emission maps obtained using the evolving particle spectra with an initial power-law index, p = 6 for the UNI-B case at $t/t_{\rm a} = 3.75$ projected on the sky plane $X^{\prime}$-$Y^{\prime}$. These maps are produced for a direction along a line of sight inclined at 20$^{\circ}$ with the jet axis at observing frequencies of $\nu/\nu_{\rm sc} \approx$ 8.2 $\times$ 10$^{4}$, 8.2 $\times$ 10$^{6}$ and 8.2 $\times$ 10$^{8}$, where $\nu_{\rm sc}$ = 122 Hz is the frequency scale. 
Middle panels: Corresponding inverse-Compton emission maps at observing energies of $E/h\nu_{\rm sc} \approx$ 7.9 $\times$ 10$^{16}$, 7.9 $\times$ 10$^{18}$, and 7.9 $\times$ 10$^{18}$, where h is the Planck's constant 
with the other model parameters being the same. The specific intensity is normalized to $I_{\rm \nu_{\rm sc}} = 7.46 \times 10^{-15}$ ergs s$^{-1}$ cm$^{-2}$ Hz$^{-1}$ str$^{-1}$ corresponding to $\gamma_{\rm max}$ = 10$^{8}$. Bottom Panels: The maps of density $\rho/\rho_0$ (left-hand panel) and pressure $P/P_0$ (right-hand panel) integrated along the same line of sight.}
\label{fig:Figure7}
\end{figure*}

Using evolving particle spectra, we choose a spectral index $p=6$ in the reference case as opposed to the A cases for which we have a spectral index p = 3 for the initial power-law distribution of non-thermal particles with initial energy bounds as $\gamma_{\rm min} = 100$ and $\gamma_{\rm max} = 10^{8}$ distributed equally in log-space using 512 bins. The ratio of the number density of the non-thermal electrons to fluid number density is kept the same as in the A cases (i.e., $\eta^{\rm NT} = 10^{-3}$). 
We use the evolving particle spectra to obtain both synchrotron and inverse-Compton emission maps at three different energies for the B cases at $t/t_{\rm a} = 3.75$ for a direction along a line of sight inclined at $\theta = 20^{\circ}$ with the jet axis in the X-Z plane. The observing frequencies for the synchrotron emission maps in scaled units are $\nu/\nu_{\rm sc} = 8.2 \times [10^4, 10^6, 10^8]$, which correspond to radio for $\nu_{\rm sc} \sim 122$ Hz. The energies for the inverse-Compton emission maps in scaled units range from $E/h\nu_{\rm sc} = 7.9 \times [10^{16}, 10^{18}, 10^{20}]$ and correspond to emission in $\gamma$-rays. The 3D distributions of pressure and density are integrated along the same line of sight. The resulting pressure and density maps along with the emission maps for the reference case UNI-B are shown in Fig.~\ref{fig:Figure7}.
A complex network of shocks evolves due to the turbulence during the non-linear phase of the KH instability. As the macro-particles encounter multiple shocks, the non-thermal electrons get re-accelerated to higher energies depending on the strength of the shocks. Consequently, the particle spectra flatten and we see bright emission features in the intensity maps at all three energies coinciding with regions at high pressure and density (see bottom panels in Fig.~\ref{fig:Figure7}). 
In particular, the synchrotron emission maps for higher frequency values (middle and right panels of the top row) show bright emission structures consistent with density and pressure maps. While the synchrotron intensity map at the lowest frequency $\nu_{1}/\nu_{\rm sc} = 8.2 \times 10^4$ is rather more diffused as expected and also shows bright emission in the center. 
Similar diffused emission is also seen for lowest IC emission energy (left middle panel), though the general features of the emission map are consistent with shocked regions depicted by pressure map. Further, at the high energy $E_{3}/h\nu_{\rm sc} = 7.9 \times 10^{20}$, the emission map shows the presence of scattered bright spots.
These findings are consistent with the results obtained by \cite{Micono1999} for the spectral index as the particles encounter multiple shocks in a 2D slab jet experiencing the KH instability. 

Further, we also examined the effects of the boundary on the qualitative behaviour of the synthetic SED. 
In particular, we increased the domain size to $ 8 R_{\rm j} \times 8 R_{\rm j} \times L_{\rm z}$ to study the influence of side boundaries on the jet dynamics and radiative properties. 
We found that the jet dynamics are slightly altered whereas the radiative properties remain the same qualitatively showing spectral hardening. 
In addition to the intensity maps, the SEDs for the B cases with two different inclination angles with static particle spectra are also discussed in the appendix \ref{app:validation}.

\begin{figure*}
\centering
\includegraphics[width = 18.0cm, scale=1.0]{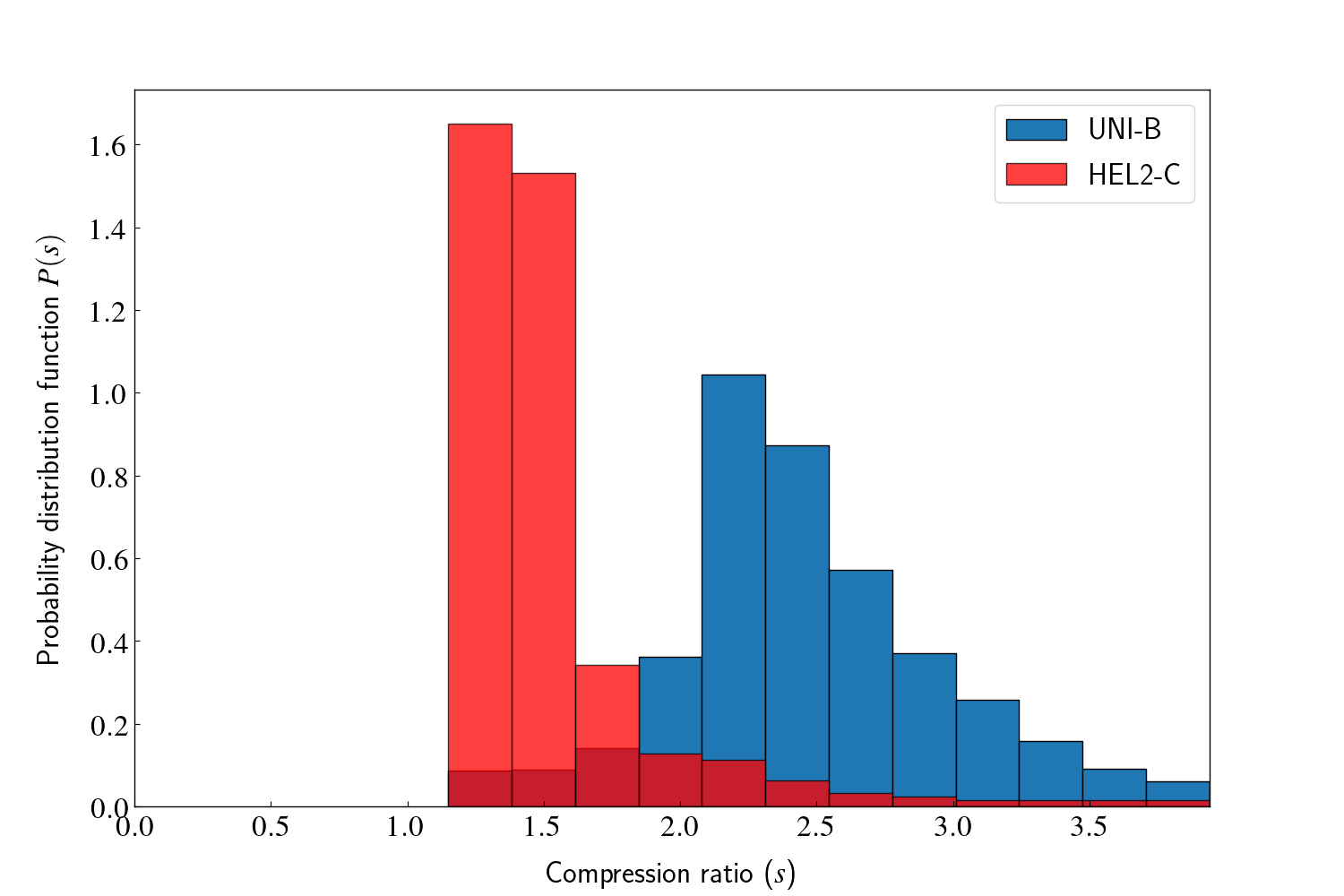}
\caption{A histogram showing the probability density distribution of the number of shocked macro-particles with compression ratio for the 'UNI-B' and 'HEL2-C' cases at $t/t_{\rm a} = 3.75$. The dark red color is due to the overlapping of the blue and red bars. }
\label{fig:Figure_extra}
\end{figure*}

A histogram of the probability density distribution function of the macro-particles with a compression ratio $s$ is shown in Fig.~\ref{fig:Figure_extra} for the reference case UNI-B (blue bars) 
at $t/t_{\rm a} = 3.75$. The area under the histogram in each bin is the probability of a macro-particle having a compression ratio between $s$ to $s + \Delta s$ where $\Delta s = 0.2$ is the bin width. 
We have about $5.5 \%$ particles in the reference case UNI-B
with a compression ratio $s > 4$. 
They typically arise when the shocks begin to steepen and cover only one or two grids zones in the transverse direction. These particles have been neglected in Fig.~\ref{fig:Figure_extra} and not accounted for calculations.
The blue histogram for the reference case UNI-B shown in Fig.~\ref{fig:Figure_extra} peaks at $s = 2.2$ which lies in the range $2.19 < s < 2.49$.

\begin{figure*}
\centering
\includegraphics[width=18.2cm]{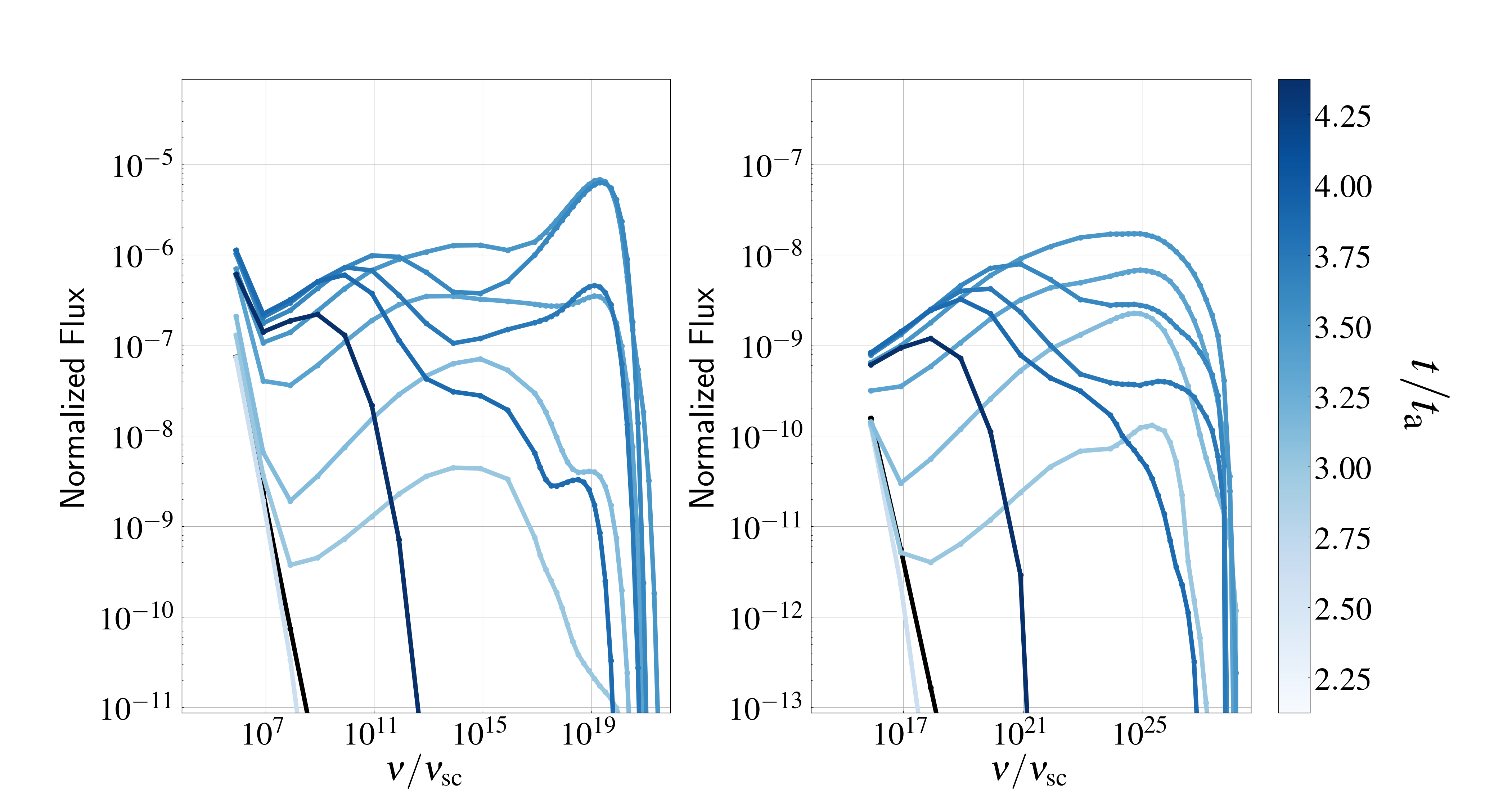}
\caption{The time evolution of the SED for model UNI-B produced using the evolving particle spectra with an initial power-law index, $p = 6$, when observed from a direction inclined at 20$^{\circ}$ with the jet axis. 
The color of the lines indicates the time normalized to Alfven time. Lines with $t/t_{\rm a} = 0.0, 2.62, 3.00, 3.12, 3.37, 3.5, 3.62, 3.75, 3.87$, and $4.37$ are shown in the left-hand panel for the synchrotron emission and corresponding IC emission is in the right-hand panel. The total integrated flux $\nu F_{\nu}$ is normalized with $\nu_{\rm sc} {F_{\nu_{\rm sc}}}$, where $\nu_{\rm sc}$ = 122 Hz is the frequency scale and ${F_{\nu_{\rm sc}}}$ =  9.38 $\times$ 10$^{-14}$ ergs s$^{-1}$ cm$^{-2}$ Hz$^{-1}$ is the flux scale corresponding to $\gamma_{\rm max}$ = 10$^{8}$. 
}
\label{fig:Figure8}
\end{figure*}

\begin{figure}
\centering
\includegraphics[width = 8.4cm]{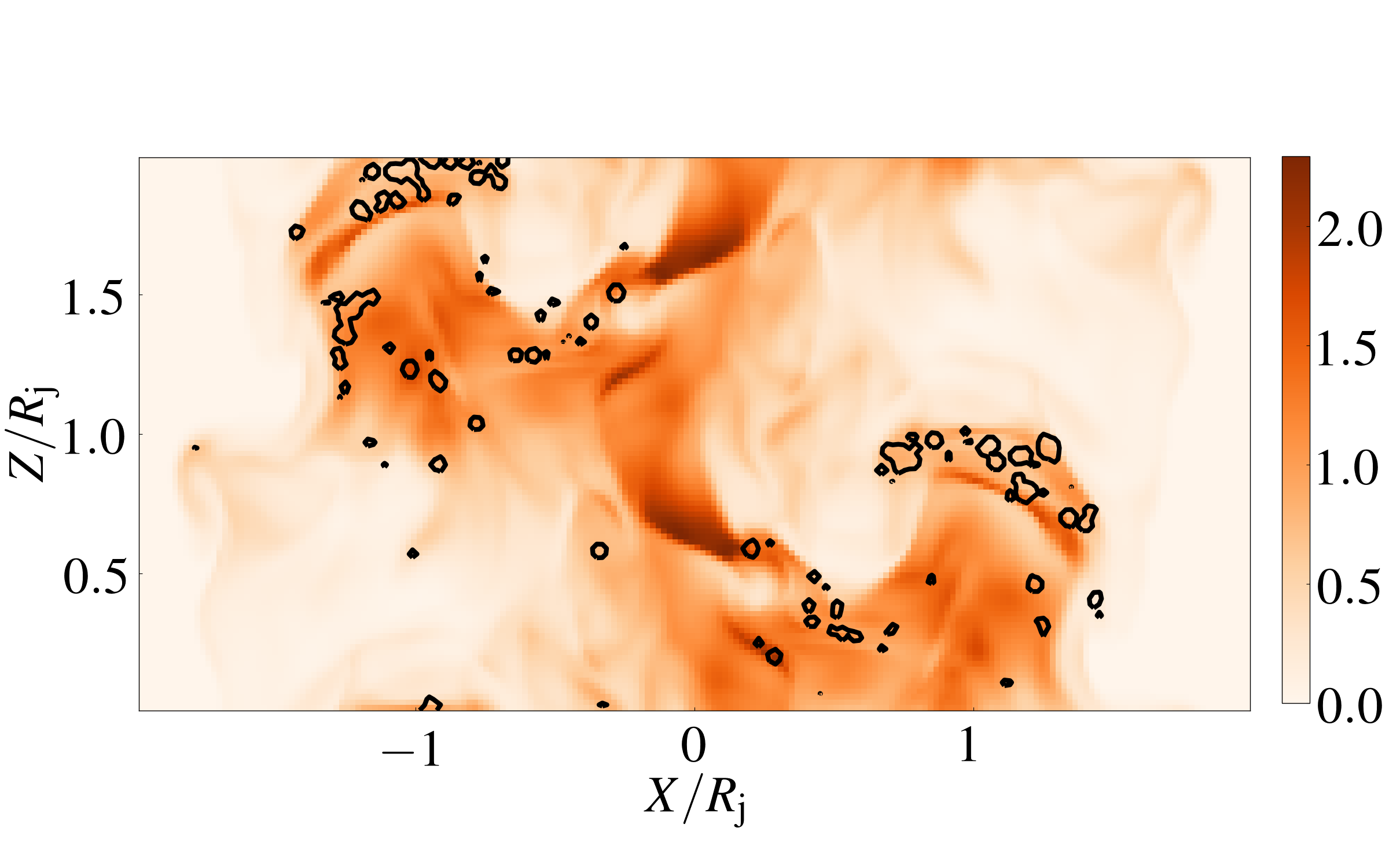}
\caption{The distribution of jet pressure $P/P_{0}$ of the reference case UNI-B in the X-Z plane at $t/t_{\rm a} = 3.75$. Over-plotted as black lines are the contours representing 1\% of the peak value of normalized synchrotron emissivity at $\nu/\nu_{\rm sc} \approx 8.2 \times 10^{14}$, where $\nu_{\rm sc}$ = 122 Hz is the frequency scale. The black lines indicate regions of X-ray spots resulting from the second population of non-thermal electrons generated in the vicinity of newly formed shocks due to shear.}
\label{fig:Figure8_new}
\end{figure}

Additionally, the effect of shocks on the jet emission can be understood by studying the temporal evolution of the SED.
The time-evolving SEDs for the reference case UNI-B are shown in Fig.~\ref{fig:Figure8}. The left panel of figure \ref{fig:Figure8} shows the evolving synchrotron SED whereas the right panel shows the corresponding IC emission. 
The initial SED is a straight line given the power-law distribution of non-thermal electrons and is shown using a black line for both the synchrotron and inverse-Compton components. This is steeper than the SED in Fig.~\ref{fig:Figure6} as the spectral index of the particle distribution has been increased to $p=6$. For the synchrotron emission, the total integrated fluxes first drop until $t/t_{\rm a} \approx 2.875$ as a result of the synchrotron cooling of the non-thermal electrons.
A sudden flattening of the SED occurs at $t/t_{\rm a} \approx 2.875$ as the turbulence results in freshly formed shocks that are captured using the evolving particle spectra. The interaction of multiple shocks accelerates the non-thermal particles to higher energies as the particle distribution spectra flatten.
Using the hybrid framework in the {\sc{PLUTO}} code, a second population of non-thermal electrons with Lorentz factors $\gamma \geq 10^{7}$ is seen.
The emergence of this second population of electrons occurs in the vicinity of freshly formed shocks and is demonstrated using synchrotron emissivity contours at $\nu/\nu_{\rm sc} \approx 10^{15}$ shown as black lines in Fig.~\ref{fig:Figure8_new}.
We observe that the maximum Lorentz factor for the spectrum associated with macro-particles in the vicinity of newly formed shocks is of the order $\gamma \approx 10^{9}$ and typical magnetic field strength of $B  = 100 \mu$\,G. This amounts to a gyro-frequency $\nu_{\rm G} = 280$Hz which implies that the critical frequency of synchrotron emission given by $\nu_{\rm c} \approx 1.5 \gamma^{2} \nu_{\rm G}$ is estimated to be of the order of $10^{21}$Hz. 
This is consistent with the fact that the peak of the bump in X-rays/$\gamma$-rays lies at $\nu/\nu_{\rm sc} \approx 10^{19}$. 
However, the shape of the SED evolves as it is a transient phenomenon. The particle acceleration due to shocks is more efficient at higher energies. However, we do see flatter spectra at lower energies as the non-thermal electrons at higher energies cool down and populate the low energy levels.

Due to the contribution from the second synchrotron component, the slope of the SED at $t/t_{\rm a} = 3.5$ 
goes directly as ${\frac{\rm 3-p}{2}}$ and equals 0.10 between the scaled frequencies $\nu/\nu_{\rm sc} = 10^{9}$ and $10^{13}$ and 0.03 between the scaled frequencies $\nu/\nu_{\rm sc} = 10^{13}$ and $10^{15}$. These slopes drop to -0.26 and -0.01 at $t/t_{\rm a} = 3.75$ for the same frequency ranges respectively due to synchrotron cooling of the newly emerged second population of non-thermal electrons. The spectral indices corresponding to these slopes at $t/t_{\rm a} = 3.75$ are $p = 3.51$ and $p = 3.02$ while the compression ratios obtained using the standard diffusive shock acceleration theory $p = 3s/(s-1) -2$ are $s = 2.19$ and $s = 2.49$ respectively indicating that the shocks are moderately strong. 

The inverse-Compton emission follows a similar trend of temporal evolution as it results from CMB photons scattering off the same non-thermal electron population which is responsible for the synchrotron emission.
\cite{Kobak2000} studied particle acceleration due to shocks using Monte Carlo simulations. \cite{Stawarz2002} derived the emission spectra for such a distribution of non-thermal electrons and found an increased emission at higher energies similar to the results shown in Fig.~\ref{fig:Figure8}.

\subsection{Comparison with other runs}\label{sec:emission_Comp}

\subsubsection{Comparison with helical jets}\label{instability_SED2}
\begin{figure*}
\centering
\includegraphics[width = 18.0cm, scale=1.0]{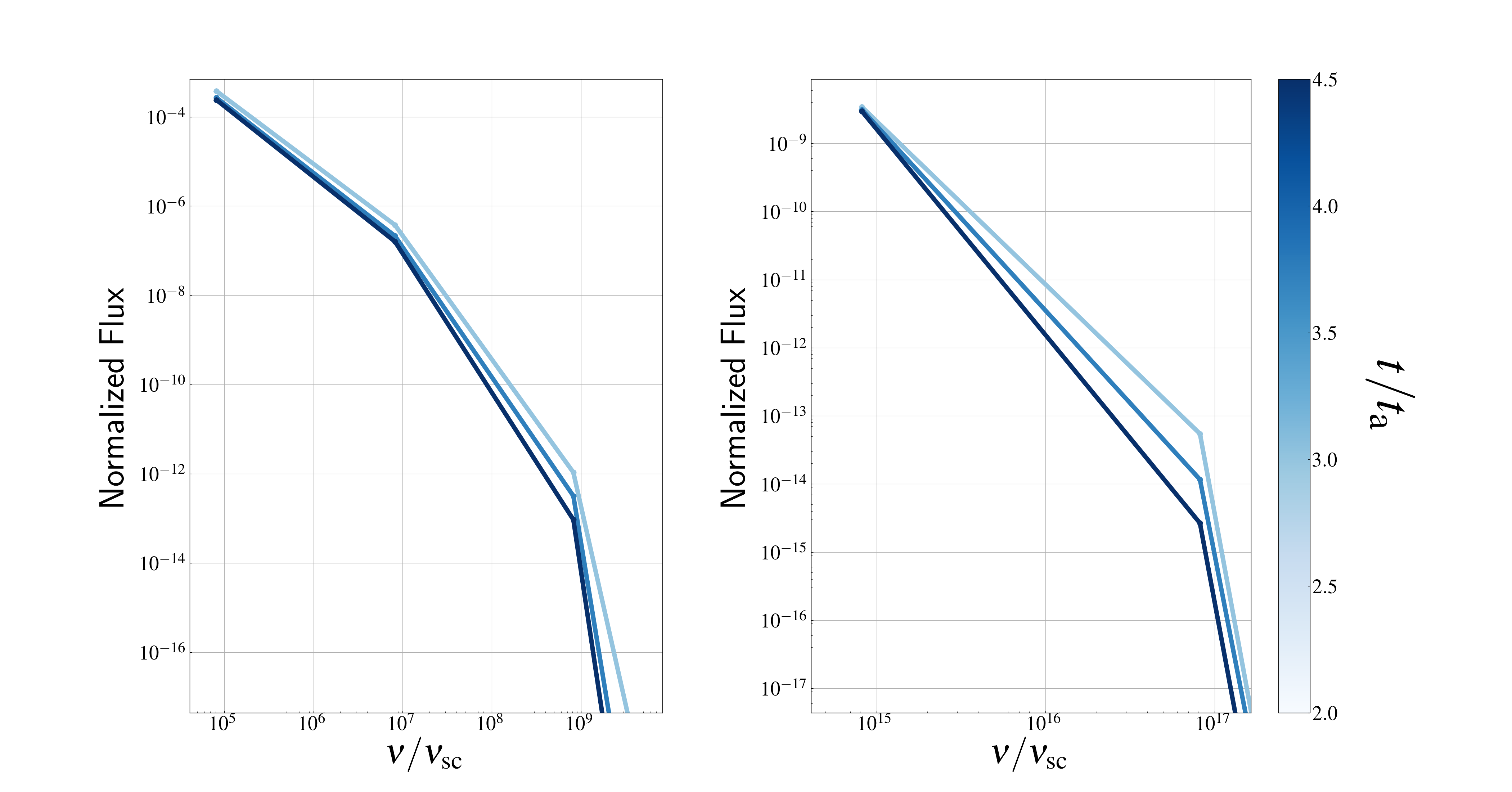}
\caption{The time evolution of the SED for model HEL2-C produced using the evolving particle spectra with an initial power-law index, $p = 6$ when observed from a direction inclined at 20$^{\circ}$ with the jet axis. 
Lines with $t/t_{\rm a} = 3.0, 3.75$, and $4.5$ are shown in the left-hand panel for the synchrotron emission and corresponding IC emission is in the right-hand panel. The frequency, flux, and time have the same normalization as in Fig.~\ref{fig:Figure8}.
}
\label{fig:Figure9}
\end{figure*}

To investigate the role of instabilities in MHD jets using radiative signatures, we compare the emission spectra of the uniform magnetic field configuration UNI-B with the helical magnetic field configurations HEL2-C obtained using the evolving particle spectra. The synchrotron and IC SED for HEL2-C case at $t/t_{\rm a} = 3.0, 3.75,$ and $4.5$ are shown in Fig.~\ref{fig:Figure9}.
In the UNI-B case, we see a flattening of the synchrotron component of the SED as a result of particle acceleration because of the freshly formed shocks which gives rise to the second population of non-thermal electrons at high energies.
As mentioned in section \ref{sec:emission_WithShocks}, the flattening of the spectra happen after $t/t_{\rm a} \approx 2.875$, when freshly formed shocks first appear. The number of shocked Lagrangian macro-particles grows rapidly with time. The probability distribution of these shocked particles represented by the blue histogram shown in Fig.~\ref{fig:Figure_extra} peaks at a compression ratio $s = 2.2$ indicating the presence of moderately strong shocks.

The jet boundary is clearly identifiable in the HEL2-B case (see figure~\ref{fig:Figure2}) as the helical magnetic field suppresses vortex formation. As a result, no shock formation occurs in the HEL2-B case. In the HEL2-C case, the perturbations do not steepen enough due to the helical magnetic field to form strong shocks despite ramping up the initial sonic Mach number. Consequently, the HEL2-C case shows the presence of weak shocks that are represented by the probability density distribution function in the red histogram in Fig.~\ref{fig:Figure_extra}. As most of the weakly shocked particles in the HEL2-C case have a compression ratio $s = 1.3$, we get $p = 3s/(s-1) -2 = 11$ for these particles making the spectrum steep (see Fig.~\ref{fig:Figure9}). The total integrated fluxes at high energies drop as a result of synchrotron cooling. Further, the HEL2 cases also have higher magnetic field strengths at the interface when compared to the UNI-B case, this as well enhances the radiative cooling observed in these runs. Negligible emission is seen beyond scaled frequency $\nu/\nu_{\rm sc} > 10^{10} $ as the particle acceleration is inefficient due to the shocks being either absent or weaker in strength in the helical jet configurations. The corresponding inverse-Compton components of the spectra have similar shapes as the same electron population is responsible for both synchrotron and IC-CMB emission. The shapes of the spectra appear to be like discontinuous curves with sharp edges as a result of an artificial effect. This is due to the fact that the SEDs have a limited resolution in the frequency domain.

In conclusion, the jet configuration with the Kelvin-Helmholtz instability alone has disruptive flow causing shock formation which results in a flatter emission spectrum while the helical jet configuration is relatively stable with weaker shocks leading to a steeper emission spectrum. 

\subsubsection{Comparison with under-dense jets}\label{under_dense_SED}

In the reference case UNI-B, no biconical shocks form near the jet axis as the density and pressure are uniform initially. For comparison, we model the jet configurations UNI-C and UNI-D that are under-dense as compared to the ambient (see Eq.~\ref{eq:Equation7}). The ambient-jet density contrast $\eta$ leads to the formation of weak biconical shocks in the UNI-D case coinciding with regions at high pressure on the jet axis as shown in the bottom-left panel in Fig.~\ref{fig:Figure3_new}.

We expect the presence of the biconical shocks to have an effect on the emission spectra. However, the emission spectra of these under-dense jets are similar to that in the reference case UNI-B. 
In the under-dense jet UNI-D, it takes a longer time for the shocks to form due to higher jet speed. 
We find that the spectral features for model UNI-D are similar to those in the reference case UNI-B (figure not shown).
It is inferred that the particle acceleration due to the weak biconical shocks is not as efficient as compared to the shocks formed at the shear surface. The minor difference in SED that arises between the UNI-B and UNI-D case particularly at low energies can be attributed to the gradual steepening of weak biconical shocks.

In order to quantify the strength of biconical shocks at $t/t_{\rm a} = 13.25$, we consider the shocked macro-particles in the cylindrical region $0 < r < a$ and the cylindrical shell $R_{\rm j} - a < r < R_{\rm j} + a$, where $a = 0.1$ is the shear layer width and $R_{\rm j} = 1.0$ is the jet radius. As the biconical shocks are weak, the number of shocked particles in the cylindrical region around the axis is less (< 100) and the majority of them have a compression ratio in the range $1.5 < s < 2.0$. The biconical shocks are weak as the density contrast $\eta$ is not large enough. In contrast, the stronger shocks at the jet boundary result in a large number (> 10$^4$) of shocked particles in the cylindrical shell and most of them have a compression ratio in the range $2.0 < s < 2.5$. The formation of strong shocks at the jet boundary is similar to what occurs in the reference case UNI-B. This is consistent with our understanding of the emission spectra in the under-dense jet UNI-D being similar to that in the reference case with minor deviation in the low energy part.

\section{Discussion}

In the present work, we have investigated the effects of MHD instabilities on the dynamics of the plasma column and their impact on synthetic non-thermal emission. In particular, we have carried out 3D simulations of a plasma column with different magnetic field configurations and shear velocity between the column and the ambient medium. 

From the dynamical evolution of the plasma column, the configuration with a dominant axial magnetic field (UNI-A) is unstable due to the KH mode. However, the presence of an azimuthal magnetic field suppresses the growth of instability and stabilizes the jet (HEL1-A and HEL2-A). We also carried out simulation runs with varying shear flow velocities that are super-sonic. The evolution of the plasma column with initial axial sonic Mach number $M_{\rm s} = 5.0$ and dominant axial magnetic fields (UNI-B case) depicted the presence of a complex network of shocks due to KHI. These shocks are typically seen near the interface between the plasma column and the ambient. In addition to these shocks, we also find the presence of weak biconical shocks for under-dense cases (UNI-C and UNI-D).

This complex network of shocks produced in the shear layer of the plasma column due to KHI is responsible for accelerating particles. From our evolving spectra analysis, we observe, local re-acceleration of electrons within the vicinity of these shocks (see figure \ref{fig:Figure8_new}). The evolution of resultant synthetic spectra is shown in figures \ref{fig:Figure8} and \ref{fig:Figure9} for the cases UNI-B and HEL2-C. On a certain evolutionary stage, the spectra show typical features corresponding to particle acceleration as multiple humps and radiative cooling as sharp cut-offs respectively. 

To quantify the amount of flux received in different frequency bands from radio to $\gamma$-rays, it is required to express the normalized values in terms of physical units. For the reference case UNI-B, we have $B_{0}$ = 44 $\mu$G at the axis at $t = 0$, which corresponds to $\nu_{\rm sc}$ = 122Hz. Further, the flux scale is ${F_{\rm \nu}}_{\rm sc}$ = 9.38 $\times 10^{-14}$ ergs s$^{-1}$ cm$^{2}$ Hz for $\gamma_{\rm max} = 10^{8}$, which is the maximum Lorentz factor set for the initial power-law spectra. With these scales, we observe that
the synthetic multi-wavelength synchrotron spectrum for the UNI-B case becomes flat in optical to X-ray band ( i.e., between $\nu = 10^{15}$ to $10^{19} $Hz) at $t/t{\rm a} \approx 3.37$ due to the formation of shocks as the result of KHI ( see left panel of figure \ref{fig:Figure8}). 
The generation of shocks in localized regions results in the re-acceleration of electrons that have a flat spectral slope consistent with the evolution of spectral index using 2D MHD runs as discussed in \cite{Micono1999}. These electrons are accelerated up to a maximum energy $\gamma = 10^9$,
forming a secondary population of energetic electrons. This exhibits a unique signature of a secondary hump in the synchrotron spectra peaking at $\nu = 10^{21} $Hz with a corresponding flux density $\nu F_{\nu} \approx 10^{-16}$ ergs s$^{-1}$ cm$^{-2}$. Additionally,
such a population of few tens to hundreds of TeV energies gives rise to TeV $\gamma$-rays ($\nu \sim 10^{25}-10^{28}$\,Hz) due to the up-scattering of CMB photons \citep{Meyer:2015, Breiding2017}.

On the other hand, due to weak shocks obtained for the HEL2-C case (see Fig.~\ref{fig:Figure_extra}), the secondary population of such high energy electrons is not formed. This results in the absence of a secondary hump in the synchrotron spectra for the HEL2-C case, instead, we observe sharp cut-offs due to effective synchrotron cooling around $\nu/\nu_{\rm sc} \approx 10^{10}$ (see Fig.~\ref{fig:Figure9}). A similar evolution of the SED is also seen for the IC emission in the HEL2-C case.

The origin of X-ray emission in the multi-wavelength SED of kilo-parsec scale AGN jets is still an open question.
\cite{Breiding2017} have proposed two possible models to explain this: synchrotron emission produced by a secondary population of higher energy electrons, and hadronic emission processes such as proton synchrotron emission or the synchrotron emission produced by a secondary population of electrons arising from photo-hadronic interactions. Our findings discussed in section \ref{sec:emission_WithShocks} indicate that the former of these models, synchrotron emission from a second electron population is a viable explanation for the high X-ray emission in the multi-wavelength SED of kilo-parsec scale AGN jets.

It should be noted that the synthetic SED obtained from this work is from a rather idealized simulation where the focus was on the portion of a jet at kpc scale and not the whole jet. Therefore the synthetic SEDs that are generated demonstrate the effects due to dynamical features that happen locally. Although these effects occur in localized regions, they are likely to happen in other regions of the jet and can be thought of as representative of a portion of the jet. Therefore, the synthetic SEDs generated can only be qualitatively compared with observations.

\section{Summary and Outlook}\label{summary}

We have carried out 3D simulations of a plasma column that represents a section of a jet at kilo-parsec scales. 
This study is aimed to understand the effects of MHD instabilities on jet dynamics and energetics that have implications for stability and emission signatures.

\begin{itemize}
   
    \item Dynamical analysis of simulation runs for A cases have been validated with those of  \cite{Baty_2002}. In particular for uniform density cases with sonic Mach number $M_{\rm s} = 1.26$, we find that the case with axial magnetic field (UNI-A) is unstable due to the KH mode and results in a disrupted and turbulent jet. On incorporating the azimuthal field, the jet column achieves stability as the growth of vortices at the shear surface is damped due to the presence of a helical field (HEL1-A and HEL2-A). (see Fig.~\ref{fig:Figure2}.)
    \item Cases with higher shear flow velocity and/or high density contrast (B, C, and D cases) show a generation of more vorticity at the shear surface and also show a presence of shocks in comparison to the A cases (see Fig.~\ref{fig:Figure3_new}). Additionally, the under-dense jet columns (C and D cases) show a presence of weak biconical shocks.
    The onset of instability in these cases happens at a later Alfven time as compared to A cases. The jet column with axial magnetic fields and high sonic Mach number is prone to KH mode instability and shows a presence of more and stronger shocks as compared to the cases with helical field structure having the same sonic Mach number.  
    \item The impact of instabilities on emission signatures has been demonstrated by generating synthetic emission maps and SEDs using static power-law spectra. We find differences in SEDs for UNI cases in comparison to ones with helical magnetic fields. These differences become more acute with a smaller viewing angle due to the dependence of synchrotron emissivity on $|\bvec{B} \times \hat {\bvec{n}}_{\rm los}| = |B| \sin \alpha $ (see Figure~\ref{fig:Figure6}).
    
    \item In order to capture localized physical effects such as particle acceleration due to shocks and radiative losses, we produced emission signatures using a more accurate model that uses evolving particle spectra. A significant difference in terms of multi-band intensity maps has been seen in comparing the static and evolving spectra models (see Fig.~\ref{fig:Figure7_new}) for the UNI-A case. This difference is purely on the account of synchrotron cooling of the non-thermal electrons as the UNI-A case with a lower axial sonic Mach number $M_{\rm s} = 1.26$ does not show any shocks. Similar features of enhanced emission at the shear surface due to compressing magnetic fields are seen in both the emission models, especially at low frequencies. 

    \item Using the evolving particle spectra for the uniform magnetic field configuration UNI-B, we see a flattening of the SED due to localized physical effects such as particle acceleration caused by the steepening of perturbations resulting in freshly formed shocks at small scales due to the KH instability. In the helical magnetic field configurations, HEL2-B and HEL2-C, the SED is relatively steeper at higher energies as the changes in jet density occur at large scales and localized shocks are either absent or weaker in strength.
    
     \item The major finding from this study is the demonstration of spectral hardening resulting in multi-peaked synthetic SED generated using the hybrid model of evolving particle spectra. The onset of KH instabilities gives rise to strong shocks at the shear surface thereby accelerating particles nearby strong shocks and generating a localized second population of high energy particles.  For our reference run, we find that the particles can be accelerated up to $\gamma_{\rm max} \sim 10^9$ and in the presence of moderate magnetic field strengths of $\sim 100 \,\mu$G, result in a transient peak at $10^{21}$Hz ($\sim 4$\,MeV) due to the synchrotron process. This can potentially explain the bright X-ray emission seen in kpc scale AGN jets.
     Another consequence of the second population of electrons is that they can produce TeV $\gamma$-rays as a result of the IC/CMB process \citep{Meyer:2015, Breiding2017}.
    
    \item For cases with an under-dense jet column, we find that the bulk of the emission arises from strong shocks formed at the shear surface, whereas, the bi-conical shocks formed near the jet axis are relatively weaker and do not contribute significantly to the hardening of spectra for the cases considered here. 
\end{itemize}

To summarise, the presence of Kelvin-Helmholtz instability alone disrupts the flow causing shock formation which results in a flatter emission spectrum whereas the inclusion of a helical magnetic field hinders the growth of Kelvin-Helmholtz instability and has a stabilizing effect on the jet making the emission spectra steep due to the absence of strong shocks. The spectral hardening due to the production of a shock accelerated localized second population of high energy electrons can provide a qualitative explanation of the bright X-ray spots typically observed in AGN jets at kilo-parsec scales.  
Another observable that can constrain the properties of AGN jets at kilo-parsec scales is the polarization. By studying the polarization properties of the jet emission, we can probe into the magnetic field structure of the jet \citep{Avachat2016}. The magnetic field structure can reveal important clues about how shocks may form and lead to particle acceleration in large scale jets. 
We aim to study the polarization properties of the jet emission in future work.

\section*{Acknowledgements}
The authors would like to thank the referee for providing valuable comments and helping to improve the manuscript significantly. 
B.V. acknowledges the financial support from the Max Planck Partner Group Award. The computations presented in this work are carried out using the facilities provided at IIT Indore and the Max Planck for Astronomy Cluster ISAAC \url{https://www.mpcdf.mpg.de/services/computing/linux/Astronomy}.

The data underlying this article will be shared on reasonable request to the corresponding author.

\bibliographystyle{aa}
\bibliography{export-bibtex}

\begin{thebibliography}{57}
\expandafter\ifx\csname natexlab\endcsname\relax\def\natexlab#1{#1}\fi

\bibitem[{{Achterberg} {et~al.}(2001){Achterberg}, {Gallant}, {Kirk}, \&
  {Guthmann}}]{Achterberg_2001}
{Achterberg}, A., {Gallant}, Y.~A., {Kirk}, J.~G., \& {Guthmann}, A.~W. 2001,
  \mnras, 328, 393

\bibitem[{{Aharonian} {et~al.}(2010){Aharonian}, {Kelner}, \&
  {Prosekin}}]{Aharonian2010}
{Aharonian}, F.~A., {Kelner}, S.~R., \& {Prosekin}, A.~Y. 2010, \prd, 82,
  043002

\bibitem[{{Appl} \& {Camenzind}(1992)}]{Appl1992}
{Appl}, S. \& {Camenzind}, M. 1992, \aap, 256, 354

\bibitem[{{Avachat} {et~al.}(2016){Avachat}, {Perlman}, {Adams}, {Cara},
  {Owen}, {Sparks}, \& {Georganopoulos}}]{Avachat2016}
{Avachat}, S.~S., {Perlman}, E.~S., {Adams}, S.~C., {et~al.} 2016, \apj, 832, 3

\bibitem[{{Baty} \& {Keppens}(2002)}]{Baty_2002}
{Baty}, H. \& {Keppens}, R. 2002, \apj, 580, 800

\bibitem[{{Birkinshaw}(1991)}]{Birkinshaw_1991}
{Birkinshaw}, M. 1991, \mnras, 252, 505

\bibitem[{{Blandford} \& {Ostriker}(1978)}]{Blandford_1978}
{Blandford}, R.~D. \& {Ostriker}, J.~P. 1978, \apjl, 221, L29

\bibitem[{{Bodo} {et~al.}(2013){Bodo}, {Mamatsashvili}, {Rossi}, \&
  {Mignone}}]{Bodo:2013}
{Bodo}, G., {Mamatsashvili}, G., {Rossi}, P., \& {Mignone}, A. 2013, \mnras,
  434, 3030

\bibitem[{{Bodo} {et~al.}(2016){Bodo}, {Mamatsashvili}, {Rossi}, \&
  {Mignone}}]{Bodo16}
{Bodo}, G., {Mamatsashvili}, G., {Rossi}, P., \& {Mignone}, A. 2016, \mnras,
  462, 3031

\bibitem[{{Bodo} {et~al.}(2019){Bodo}, {Mamatsashvili}, {Rossi}, \&
  {Mignone}}]{Bodo19}
{Bodo}, G., {Mamatsashvili}, G., {Rossi}, P., \& {Mignone}, A. 2019, \mnras,
  485, 2909

\bibitem[{{Bodo} {et~al.}(1989){Bodo}, {Rosner}, {Ferrari}, \&
  {Knobloch}}]{Bodo_1989}
{Bodo}, G., {Rosner}, R., {Ferrari}, A., \& {Knobloch}, E. 1989, \apj, 341, 631

\bibitem[{{Bodo} {et~al.}(1996){Bodo}, {Rosner}, {Ferrari}, \&
  {Knobloch}}]{Bodo_1996}
{Bodo}, G., {Rosner}, R., {Ferrari}, A., \& {Knobloch}, E. 1996, \apj, 470, 797

\bibitem[{{Breiding} {et~al.}(2017){Breiding}, {Meyer}, {Georganopoulos},
  {Keenan}, {DeNigris}, \& {Hewitt}}]{Breiding2017}
{Breiding}, P., {Meyer}, E.~T., {Georganopoulos}, M., {et~al.} 2017, \apj, 849,
  95

\bibitem[{{Drury}(1983)}]{Drury_1983}
{Drury}, L.~O. 1983, Reports on Progress in Physics, 46, 973

\bibitem[{{Frank} {et~al.}(1996){Frank}, {Jones}, {Ryu}, \&
  {Gaalaas}}]{Frank:1996}
{Frank}, A., {Jones}, T.~W., {Ryu}, D., \& {Gaalaas}, J.~B. 1996, \apj, 460,
  777

\bibitem[{{Fromm} {et~al.}(2019){Fromm}, {Younsi}, {Baczko}, {Mizuno}, {Porth},
  {Perucho}, {Olivares}, {Nathanail}, {Angelakis}, {Ros}, {Zensus}, \&
  {Rezzolla}}]{Fromm:2019}
{Fromm}, C.~M., {Younsi}, Z., {Baczko}, A., {et~al.} 2019, \aap, 629, A4

\bibitem[{{H.~E.~S.~S. Collaboration} {et~al.}(2020){H.~E.~S.~S.
  Collaboration}, {Abdalla}, {Adam}, {Aharonian}, {Ait Benkhali},
  {Ang{\"u}ner}, {Arakawa}, {Arcaro}, {Armand}, {Ashkar}, {Backes}, {Barbosa
  Martins}, {Barnard}, {Becherini}, {Berge}, {Bernl{\"o}hr}, {Blackwell},
  {B{\"o}ttcher}, {Boisson}, {Bolmont}, {Bonnefoy}, {Bregeon}, {Breuhaus},
  {Brun}, {Brun}, {Bryan}, {B{\"u}chele}, {Bulik}, {Bylund}, {Capasso},
  {Caroff}, {Carosi}, {Casanova}, {Cerruti}, {Chand}, {Chandra}, {Chen},
  {Colafrancesco}, {Cury{\l}o}, {Davids}, {Deil}, {Devin}, {deWilt}, {Dirson},
  {Djannati-Ata{\"\i}}, {Dmytriiev}, {Donath}, {Doroshenko}, {Drury}, {Dyks},
  {Egberts}, {Emery}, {Ernenwein}, {Eschbach}, {Feijen}, {Fegan}, {Fiasson},
  {Fontaine}, {Funk}, {F{\"u}{\ss}ling}, {Gabici}, {Gallant}, {Gat{\'e}},
  {Giavitto}, {Glawion}, {Glicenstein}, {Gottschall}, {Grondin}, {Hahn},
  {Haupt}, {Heinzelmann}, {Henri}, {Hermann}, {Hinton}, {Hofmann}, {Hoischen},
  {Holch}, {Holler}, {Horns}, {Huber}, {Iwasaki}, {Jamrozy}, {Jankowsky},
  {Jankowsky}, {Jardin-Blicq}, {Jung-Richardt}, {Kastendieck},
  {Katarzy{\'n}ski}, {Katsuragawa}, {Katz}, {Khangulyan}, {Kh{\'e}lifi},
  {King}, {Klepser}, {Klu{\'z}niak}, {Komin}, {Kosack}, {Kostunin}, {Kraus},
  {Lamanna}, {Lau}, {Lemi{\`e}re}, {Lemoine-Goumard}, {Lenain}, {Leser},
  {Levy}, {Lohse}, {Lypova}, {Mackey}, {Majumdar}, {Malyshev}, {Marandon},
  {Marcowith}, {Mares}, {Mariaud}, {Mart{\'\i}-Devesa}, {Marx}, {Maurin},
  {Meintjes}, {Mitchell}, {Moderski}, {Mohamed}, {Mohrmann}, {Moore}, {Moulin},
  {Muller}, {Murach}, {Nakashima}, {de Naurois}, {Ndiyavala}, {Niederwanger},
  {Niemiec}, {Oakes}, {O'Brien}, {Odaka}, {Ohm}, {de Ona Wilhelmi},
  {Ostrowski}, {Oya}, {Panter}, {Parsons}, {Perennes}, {Petrucci}, {Peyaud},
  {Piel}, {Pita}, {Poireau}, {Priyana Noel}, {Prokhorov}, {Prokoph},
  {P{\"u}hlhofer}, {Punch}, {Quirrenbach}, {Raab}, {Rauth}, {Reimer}, {Reimer},
  {Remy}, {Renaud}, {Rieger}, {Rinchiuso}, {Romoli}, {Rowell}, {Rudak},
  {Ruiz-Velasco}, {Sahakian}, {Saito}, {Sanchez}, {Santangelo}, {Sasaki},
  {Schlickeiser}, {Sch{\"u}ssler}, {Schulz}, {Schutte}, {Schwanke},
  {Schwemmer}, {Seglar-Arroyo}, {Senniappan}, {Seyffert}, {Shafi},
  {Shiningayamwe}, {Simoni}, {Sinha}, {Sol}, {Specovius}, {Spir-Jacob},
  {Stawarz}, {Steenkamp}, {Stegmann}, {Steppa}, {Takahashi}, {Tavernier},
  {Taylor}, {Terrier}, {Tiziani}, {Tluczykont}, {Trichard}, {Tsirou}, {Tsuji},
  {Tuffs}, {Uchiyama}, {van der Walt}, {van Eldik}, {van Rensburg}, {van
  Soelen}, {Vasileiadis}, {Veh}, {Venter}, {Vincent}, {Vink}, {Voisin},
  {V{\"o}lk}, {Vuillaume}, {Wadiasingh}, {Wagner}, {White}, {Wierzcholska},
  {Yang}, {Yoneda}, {Zacharias}, {Zanin}, {Zdziarski}, {Zech}, {Ziegler},
  {Zorn}, \& {{\.Z}ywucka}}]{Hess_2020}
{H.~E.~S.~S. Collaboration}, {Abdalla}, H., {Adam}, R., {et~al.} 2020, \nat,
  582, 356

\bibitem[{Hardee(2008)}]{Hardee2008}
Hardee, P. 2008, Journal of Physics: Conference Series, 131, 012052

\bibitem[{{Hardee}(2006)}]{2006AIPC..856...57H}
{Hardee}, P.~E. 2006, in American Institute of Physics Conference Series, Vol.
  856, Relativistic Jets: The Common Physics of AGN, Microquasars, and
  Gamma-Ray Bursts, ed. P.~A. {Hughes} \& J.~N. {Bregman}, 57--77

\bibitem[{{Hardee} \& {Clarke}(1992)}]{Hardee_1992}
{Hardee}, P.~E. \& {Clarke}, D.~A. 1992, \apjl, 400, L9

\bibitem[{{Huber} {et~al.}(2021){Huber}, {Kissmann}, {Reimer}, \&
  {Reimer}}]{Huber:2021}
{Huber}, D., {Kissmann}, R., {Reimer}, A., \& {Reimer}, O. 2021, \aap, 646, A91

\bibitem[{{Istomin} \& {Pariev}(1994)}]{Istomin94}
{Istomin}, Y.~N. \& {Pariev}, V.~I. 1994, \mnras, 267, 629

\bibitem[{{Istomin} \& {Pariev}(1996)}]{Istomin96}
{Istomin}, Y.~N. \& {Pariev}, V.~I. 1996, \mnras, 281, 1

\bibitem[{{Jones} {et~al.}(1994){Jones}, {Kang}, \& {Tregillis}}]{Jones1994}
{Jones}, T.~W., {Kang}, H., \& {Tregillis}, I.~L. 1994, \apj, 432, 194

\bibitem[{{Kersal{\'e}} {et~al.}(2000){Kersal{\'e}}, {Longaretti}, \&
  {Pelletier}}]{Kersale2000}
{Kersal{\'e}}, E., {Longaretti}, P.~Y., \& {Pelletier}, G. 2000, \aap, 363,
  1166

\bibitem[{{Khangulyan} {et~al.}(2014){Khangulyan}, {Aharonian}, \&
  {Kelner}}]{Khangulyan2014}
{Khangulyan}, D., {Aharonian}, F.~A., \& {Kelner}, S.~R. 2014, \apj, 783, 100

\bibitem[{{Kim} {et~al.}(2016){Kim}, {Balsara}, {Lyutikov}, \&
  {Komissarov}}]{Kim16}
{Kim}, J., {Balsara}, D.~S., {Lyutikov}, M., \& {Komissarov}, S.~S. 2016,
  \mnras, 461, 728

\bibitem[{{Kim} {et~al.}(2017){Kim}, {Balsara}, {Lyutikov}, \&
  {Komissarov}}]{Kim17}
{Kim}, J., {Balsara}, D.~S., {Lyutikov}, M., \& {Komissarov}, S.~S. 2017,
  \mnras, 467, 4647

\bibitem[{{Kim} {et~al.}(2018){Kim}, {Balsara}, {Lyutikov}, \&
  {Komissarov}}]{Kim18}
{Kim}, J., {Balsara}, D.~S., {Lyutikov}, M., \& {Komissarov}, S.~S. 2018,
  \mnras, 474, 3954

\bibitem[{{Kim} {et~al.}(2015){Kim}, {Balsara}, {Lyutikov}, {Komissarov},
  {George}, \& {Siddireddy}}]{Kim15}
{Kim}, J., {Balsara}, D.~S., {Lyutikov}, M., {et~al.} 2015, \mnras, 450, 982

\bibitem[{{Kirk} {et~al.}(2000){Kirk}, {Guthmann}, {Gallant}, \&
  {Achterberg}}]{Kirk_2000}
{Kirk}, J.~G., {Guthmann}, A.~W., {Gallant}, Y.~A., \& {Achterberg}, A. 2000,
  \apj, 542, 235

\bibitem[{{Kobak} \& {Ostrowski}(2000)}]{Kobak2000}
{Kobak}, T. \& {Ostrowski}, M. 2000, \mnras, 317, 973

\bibitem[{{Laing} \& {Bridle}(2014)}]{Laing2014}
{Laing}, R.~A. \& {Bridle}, A.~H. 2014, \mnras, 437, 3405

\bibitem[{{Longair}(2011)}]{Longair2011}
{Longair}, M.~S. 2011, {High Energy Astrophysics}

\bibitem[{{Malagoli} {et~al.}(1996){Malagoli}, {Bodo}, \&
  {Rosner}}]{Malagoli1996}
{Malagoli}, A., {Bodo}, G., \& {Rosner}, R. 1996, \apj, 456, 708

\bibitem[{{Massaglia} {et~al.}(2016){Massaglia}, {Bodo}, {Rossi}, {Capetti}, \&
  {Mignone}}]{2016A&A...596A..12M}
{Massaglia}, S., {Bodo}, G., {Rossi}, P., {Capetti}, S., \& {Mignone}, A. 2016,
  \aap, 596, A12

\bibitem[{{Meyer} {et~al.}(2015){Meyer}, {Georganopoulos}, {Sparks}, {Perlman},
  {van der Marel}, {Anderson}, {Sohn}, {Biretta}, {Norman}, \&
  {Chiaberge}}]{Meyer:2015}
{Meyer}, E.~T., {Georganopoulos}, M., {Sparks}, W.~B., {et~al.} 2015, \nat,
  521, 495

\bibitem[{{Micono} {et~al.}(1999){Micono}, {Zurlo}, {Massaglia}, {Ferrari}, \&
  {Melrose}}]{Micono1999}
{Micono}, M., {Zurlo}, N., {Massaglia}, S., {Ferrari}, A., \& {Melrose}, D.~B.
  1999, \aap, 349, 323

\bibitem[{{Mignone} {et~al.}(2007){Mignone}, {Bodo}, {Massaglia}, {Matsakos},
  {Tesileanu}, {Zanni}, \& {Ferrari}}]{Mignone_2007}
{Mignone}, A., {Bodo}, G., {Massaglia}, S., {et~al.} 2007, \apjs, 170, 228

\bibitem[{{Mimica} {et~al.}(2013){Mimica}, {Aloy}, {Rueda-Becerril}, {Tabik},
  \& {Aloy}}]{2013JPhCS.454a2001M}
{Mimica}, P., {Aloy}, M.~A., {Rueda-Becerril}, J.~M., {Tabik}, S., \& {Aloy},
  C. 2013, in Journal of Physics Conference Series, Vol. 454, Journal of
  Physics Conference Series, 012001

\bibitem[{{Mizuno} {et~al.}(2007){Mizuno}, {Hardee}, \&
  {Nishikawa}}]{2007ApJ...662..835M}
{Mizuno}, Y., {Hardee}, P., \& {Nishikawa}, K.-I. 2007, \apj, 662, 835

\bibitem[{{Mukherjee} {et~al.}(2020){Mukherjee}, {Bodo}, {Mignone}, {Rossi}, \&
  {Vaidya}}]{Mukherjee_2020}
{Mukherjee}, D., {Bodo}, G., {Mignone}, A., {Rossi}, P., \& {Vaidya}, B. 2020,
  \mnras, 499, 681

\bibitem[{{Pandya} {et~al.}(2016){Pandya}, {Zhang}, {Chandra}, \&
  {Gammie}}]{Pandya2016}
{Pandya}, A., {Zhang}, Z., {Chandra}, M., \& {Gammie}, C.~F. 2016, \apj, 822,
  34

\bibitem[{{Perucho} {et~al.}(2004){Perucho}, {Hanasz}, {Mart{\'\i}}, \&
  {Sol}}]{Perucho_2004}
{Perucho}, M., {Hanasz}, M., {Mart{\'\i}}, J.~M., \& {Sol}, H. 2004, \aap, 427,
  415

\bibitem[{{Perucho} {et~al.}(2010){Perucho}, {Mart{\'\i}}, {Cela}, {Hanasz},
  {de La Cruz}, \& {Rubio}}]{Perucho_2010}
{Perucho}, M., {Mart{\'\i}}, J.~M., {Cela}, J.~M., {et~al.} 2010, \aap, 519,
  A41

\bibitem[{{Petruk}(2009)}]{Petruk2009}
{Petruk}, O. 2009, \aap, 499, 643

\bibitem[{{Rossi} {et~al.}(2008){Rossi}, {Mignone}, {Bodo}, {Massaglia}, \&
  {Ferrari}}]{Rossi_2008}
{Rossi}, P., {Mignone}, A., {Bodo}, G., {Massaglia}, S., \& {Ferrari}, A. 2008,
  \aap, 488, 795

\bibitem[{{Rybicki} \& {Lightman}(1986)}]{Rybiki1986}
{Rybicki}, G.~B. \& {Lightman}, A.~P. 1986, {Radiative Processes in
  Astrophysics}

\bibitem[{{Ryu} {et~al.}(2000){Ryu}, {Jones}, \& {Frank}}]{Ryu2000}
{Ryu}, D., {Jones}, T.~W., \& {Frank}, A. 2000, \apj, 545, 475

\bibitem[{{Schlickeiser} \& {Ruppel}(2010)}]{Schlickeiser_2010}
{Schlickeiser}, R. \& {Ruppel}, J. 2010, New Journal of Physics, 12, 033044

\bibitem[{{Schwartz}(1998)}]{Schwartz_1998}
{Schwartz}, S.~J. 1998, ISSI Scientific Reports Series, 1, 249

\bibitem[{{Stawarz} \& {Ostrowski}(2002)}]{Stawarz2002}
{Stawarz}, {\L}. \& {Ostrowski}, M. 2002, \apj, 578, 763

\bibitem[{{Tregillis} {et~al.}(2001){Tregillis}, {Jones}, \&
  {Ryu}}]{Tregillis:2001}
{Tregillis}, I.~L., {Jones}, T.~W., \& {Ryu}, D. 2001, \apj, 557, 475

\bibitem[{{Urpin}(2002)}]{Urpin_2002}
{Urpin}, V. 2002, \aap, 385, 14

\bibitem[{{Vaidya} {et~al.}(2018){Vaidya}, {Mignone}, {Bodo}, {Rossi}, \&
  {Massaglia}}]{Vaidya_2018}
{Vaidya}, B., {Mignone}, A., {Bodo}, G., {Rossi}, P., \& {Massaglia}, S. 2018,
  \apj, 865, 144

\bibitem[{{Winner} {et~al.}(2019){Winner}, {Pfrommer}, {Girichidis}, \&
  {Pakmor}}]{Winner_2019}
{Winner}, G., {Pfrommer}, C., {Girichidis}, P., \& {Pakmor}, R. 2019, \mnras,
  488, 2235

\bibitem[{{Zabalza}(2015)}]{Zabalza2015}
{Zabalza}, V. 2015, in International Cosmic Ray Conference, Vol.~34, 34th
  International Cosmic Ray Conference (ICRC2015), 922

\end{thebibliography}

\appendix

\section{Static particle spectra approach : Formulations and Validation}
\label{app:validation}

In this section, we describe in detail the formulations utilized for the static particle spectra approach (see Section~\ref{fixed_spectra}) along with its validation. Further, we also discuss the effect of the line of sight variation on the shape of the spectral distribution obtained with fixed particle spectra. 

The synchrotron emissivity (\ref{eq:Equation16}) can be expressed in the form given by \cite{Pandya2016},
\begin{equation}
\begin{split}
    J_{\rm syn}(\nu,\hat {\bvec{n}}_{\rm los}) = \frac{\nu_{\rm G} \sin \alpha (p-1) 3^{\rm p/2}}{\gamma_{\rm min}^{\rm 1-p} - \gamma_{\rm max}^{\rm 1-p}}\left( \frac{\nu}{\nu_{\rm G} \sin\alpha} \right)^{\frac{\rm 1 - p}{2}} \frac{n_{\rm e}^{\rm NT} e^{2}}{c 2^{\frac{\rm p+3}{2}}}  \\ \times \int_{x_{\rm 1}}^{x_{\rm 2}} F(x) x^{\frac{\rm p-3}{2}} dx,
\label{eq:Equation17}
\end{split}
\end{equation}
where $\alpha$ is the angle between the line of sight vector $\hat {\bvec{n}}_{\rm los}$ and the magnetic field vector $\bvec{B}$,
and $\nu_{\rm G} = eB/2\pi m_{\rm e}c$ is the gyrofrequency of an electron \citep{Longair2011}.
The function $F(x)$ is the modified Bessel function integral of the order $5/3$,
\begin{equation}
    F(x) = x \int_{\rm x}^{\infty} K_{5/3} (\eta) d\eta,
\end{equation}
\\where $x = \frac{\nu}{\nu_{\rm c}}$ and $\nu_{\rm c}$ is the critical frequency of synchrotron emission for a single electron given by
\begin{equation}
    \nu_{\rm c} = \frac{3}{2} \gamma^{2} \nu_{\rm G} \sin\alpha.
\end{equation}

Emissivity due to inverse-Compton scattering from CMB can be expressed in the form given by \cite{Rybiki1986},
\begin{equation}
\begin{split}
     J_{\rm IC}(\nu) = \frac{3 c h \sigma_{\rm T} n_{\rm e}^{\rm NT} (p-1) 2^{\rm p-2}}{4 \pi \left(\gamma_{\rm min}^{\rm 1-p} - \gamma_{\rm max}^{\rm 1-p}\right){\nu}^{\frac{\rm p-1}{2}}} \int \hspace{-1mm} d{\nu_{\rm 1}} \hspace{1mm}\nu_{\rm 1}^{\frac{\rm p-3}{2}} \epsilon(\nu_{\rm 1})\\ \int_{x_{\rm 1}}^{x_{2}}\hspace{-1mm} dx\hspace{1mm} x^{\frac{\rm p-1}{2}} f(x),
\label{eq:Equation20}
\end{split}
\end{equation}
where $\sigma_{\rm T}$ is the Thomson scattering cross-section, and $\epsilon(\nu_{\rm 1})$ is the black-body distribution of the seed photons from the Cosmic Microwave Background given by the expression
\begin{equation}
    \epsilon(\nu_{\rm 1}) = \frac{8 \pi h}{c^{3}} \, \frac{\nu_{\rm 1}^{3}}{\exp \left(\frac{h \nu_{\rm 1}}{KT_{\rm CMB}}\right) - 1},
\end{equation}
and the function $f (x)$ is given by
\begin{equation}
    f (x) = 2x \ln(x) + 1 + x - 2 x^{2}, {\hspace{10mm}}  0 {\hspace{2mm}}< \hspace{2mm}x \hspace{2mm}<\hspace{2mm}1
\end{equation}
where $x =\frac{ \nu}{4 \gamma^{2} \nu_{\rm 1}}$.
\\ For low energy electrons with Lorentz factor $\gamma << \gamma_{\rm k}$, IC losses will be in the Thomson limit, but for high energy electrons with $\gamma >> \gamma_{\rm k}$, the IC losses occur in the extreme Klein-Nishina limit where $\gamma_{\rm k} = \frac{0.53 m_{\rm e} c^{2}}{K_{\rm B} T} = 10^{9}$ is the critical Klein-Nishina Lorentz factor for a CMB photon with $T = T_{\rm CMB} = 2.73K$ \citep{Schlickeiser_2010,Petruk2009}.

\begin{figure}
\centering
\includegraphics[scale=0.25]{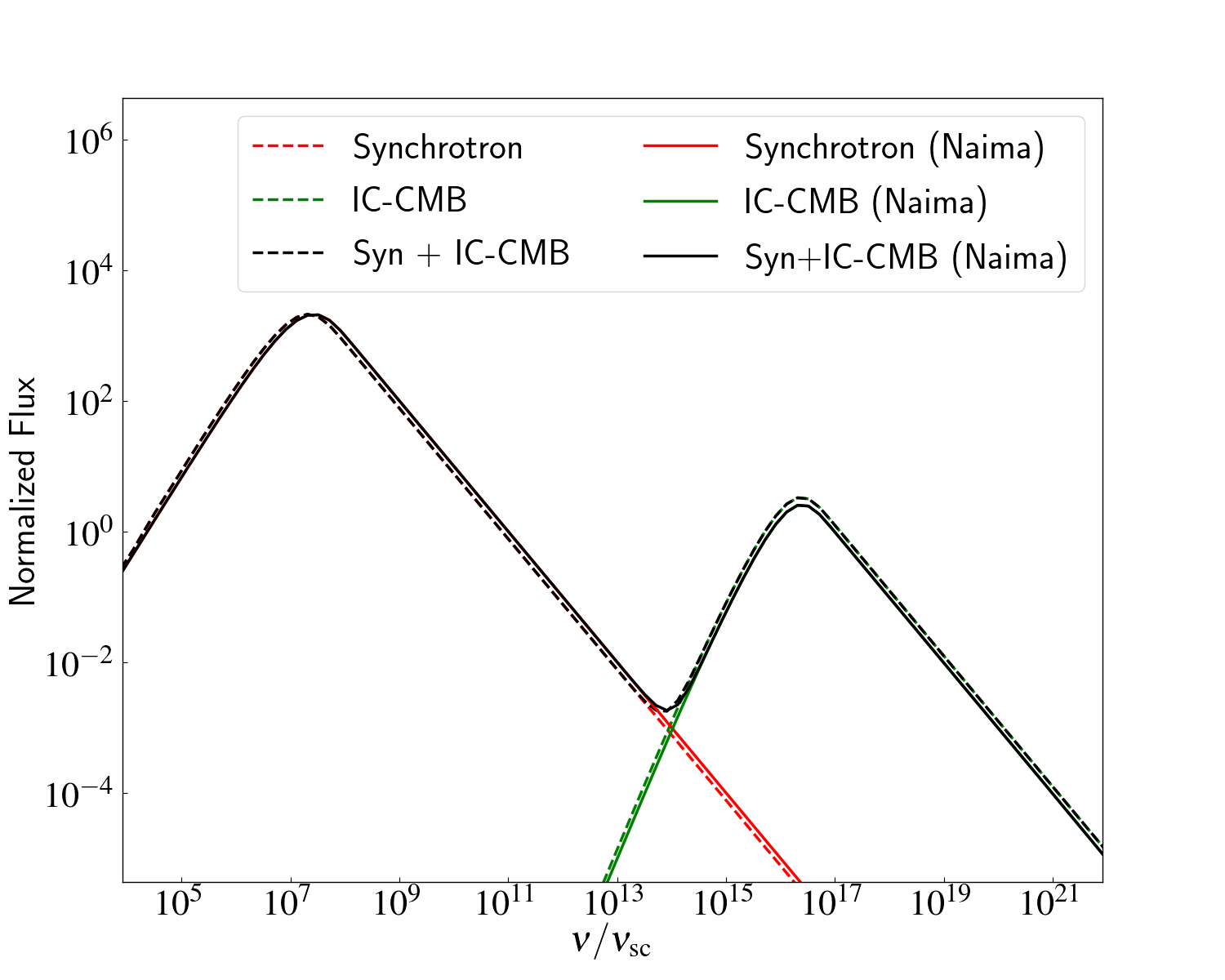}
\caption{Comparison between SED of synchrotron and inverse-Compton emission with power-law index, p = 5 obtained using the static particle spectra for a single grid cell and one-zone model {\sc{NAIMA}} keeping all other parameters the same. The flux is normalized with $\nu_{\rm sc} {F_{\nu_{\rm sc}}}$, where $\nu_{\rm sc}$ = 122 Hz is the frequency scale and ${F_{\nu_{\rm sc}}}$ =  9.38 $\times$ 10$^{-8}$ ergs s$^{-1}$ cm$^{-2}$ Hz$^{-1}$ is the flux scale corresponding to $\gamma_{\rm max}$ = 10$^{6}$.}
\label{fig:Figure4}
\end{figure}

We validate the above formulations used in our post-processing tool with static particle spectra for a single grid-cell with an open-source one zone model {\sc{NAIMA}} \citep{Zabalza2015}.
We assigned a particular set of free parameters to model the synchrotron and IC emission from the jet at kpc-scales. 
The parameters are minimum electron energy $E_{0} = 1TeV$, power-law index $p = 5$, the distance between the blob and observer $D = 1.5kpc$, magnetic field strength $ B = 100 \mu\,$G, and the ratio of the number density of non-thermal electrons to fluid number density $\eta^{\rm NT} = 0.01$. The same set of parameters are inputted to the {\sc{NAIMA}} code for obtaining the SED. 
A comparison of the SED obtained using the static particle spectra for a single grid-cell with those produced using the {\sc{NAIMA}} code is shown in Fig.~\ref{fig:Figure4}. The SEDs from both these codes show a decent match across the electromagnetic spectrum ranging from frequencies between $1$MHz in radio to $1$YHz in $\gamma$-rays. This clearly validates the equations adopted for the static particle spectra approach. While the {\sc{NAIMA}} code uses analytical approximations for computing both the synchrotron \citep{Aharonian2010} and IC \citep{Khangulyan2014} emissivities, our approach computes the exact integrals in equations~\ref{eq:Equation17} and~\ref{eq:Equation20} numerically for more accurate results.

We also study the direction dependence of the static particle spectra approach. 
For understanding this, we obtain the SEDs for the B cases with two different inclination angles. The input parameters for the static particle spectra are the same as considered for the intensity maps (as given in section \ref{sec:emission_NoShocks}). The sharp cut-offs at both ends of the spectrum are because of the energy limits of the emitting electrons whereas the slope of the flat portion of the SED is zero for our chosen value of the spectral index, $p=3$ as it goes as $(\rm 3-p)/2$.
For a direction along a line of sight that is close to the axial direction with $\theta = 20^{\circ}$, the total integrated flux in the UNI-B case normalized to $\nu_{\rm sc} F_{\nu_{\rm sc}}$ is of the order $10^{-11}$ which is nearly two orders of magnitude lower than the normalized flux values in the HEL1-B and HEL2-B cases that are of the order $10^{-9}$.
The fluxes for all the B cases are of the same order for a direction along a line of sight that is highly inclined with the jet axis ($\theta= 50^{\circ}$) in the X-Z plane. The increased synchrotron emission for the UNI-B case at higher inclination angles of the line of sight with the jet axis can be attributed to the orientation of the line of sight vector $\bvec{\hat n}_{\rm los}$ with the magnetic field vector $\bvec{B}$. 

\begin{figure*}
\centering
\includegraphics[width = 18.0 cm]{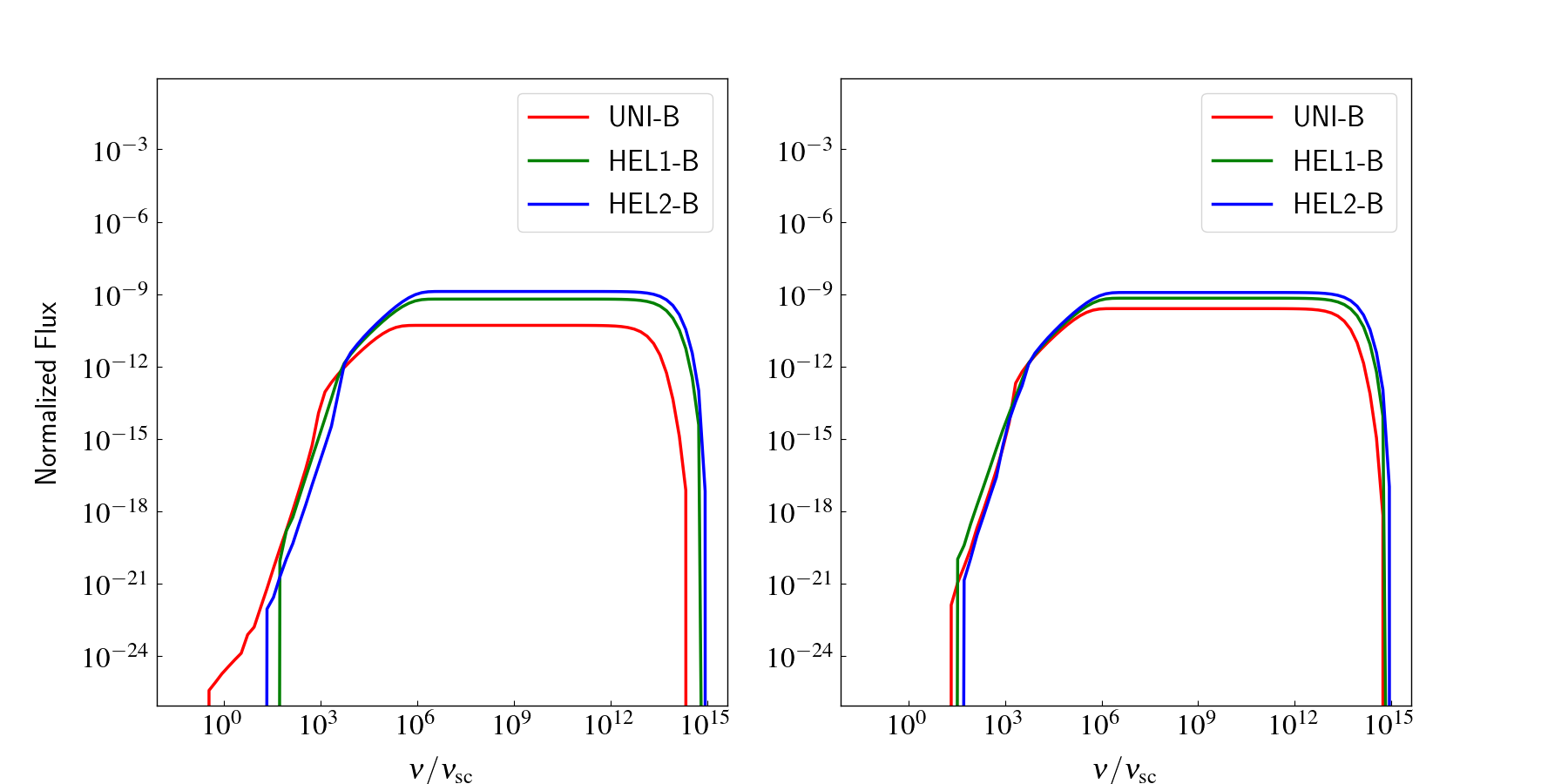}
\caption{
The SED of synchrotron emission produced using static particle spectra with power-law index, p = 3 for the B cases with $M_{\rm s} = 5.00$ at $t/t_{\rm a}$ = 2.25 when seen from a direction inclined at 20$^{\circ}$ (left-hand panel) and 50$^{\circ}$ (right-hand panel) with the jet axis. The frequency and flux have the same normalization as in Fig.~\ref{fig:Figure4}.
}
\label{fig:Figure6}
\end{figure*}

The synchrotron emissivity given by equation~\ref{eq:Equation17} goes directly as $| \bvec{B} \times \bvec{\hat n}_{\rm los}| \propto |B|\sin\alpha$ where $\alpha$ is the angle between the line of sight vector $\bvec{\hat n}_{\rm los}$ and the magnetic field vector $\bvec{B}$. In the UNI-B case, as the axial magnetic field dominates the transverse magnetic field in most of the grid zones, we get $\theta \approx \alpha$. For a line of sight close to the jet axis ($\theta = 20^{\circ}$), the contribution to the total integrated flux from most of the grid zones will be less as $\sin\alpha$ has smaller values for small $\alpha$. However, in the HEL1-B and HEL2-B cases, the azimuthal magnetic field $B_{\rm \phi}$ which is nearly perpendicular to a line of sight close to the jet axis contributes significantly to the synchrotron emission as $\sin\alpha \to 1$ for $\alpha \to 90^{\circ}$. This explains the higher total integrated flux levels in the HEL1-B and HEL2-B cases for $\theta = 20^{\circ}$. For $\theta = 50^{\circ}$, on the other hand, the axial magnetic field in the UNI-B case will now have a component that is perpendicular to the line of sight which contributes to the synchrotron emission of the jet. This results in fluxes of roughly the same order for the B cases as shown in the right-hand panel of Fig.~\ref{fig:Figure6}.

\end{document}